\documentclass[12pt]{article}

\usepackage{indentfirst,csquotes}
\usepackage{natbib}
\usepackage{color}
\usepackage{hyperref}
\usepackage{colortbl}
\usepackage[top=2.54cm, bottom=2.54cm, left=3.17cm, right=2.54cm]{geometry}
\usepackage{amssymb,amsthm,amsmath}
\usepackage{xcolor,paralist,hyperref,titlesec,fancyhdr,etoolbox}

\usepackage{authblk} 
\newcommand{\keywords}[1]{\textbf{Keywords:} #1}

\usepackage{xr}
\usepackage{setspace}

\usepackage{amsmath, amssymb, url, multirow, booktabs, verbatim, bm, threeparttable, natbib, epsfig, lscape, mathrsfs,float}
\usepackage{xcolor}   

\usepackage{threeparttable}
\usepackage{graphicx} 
\usepackage{subcaption} 
\PassOptionsToPackage{table}{xcolor}  

\usepackage{algorithm} 
\usepackage{algpseudocode}

\usepackage{enumitem,cleveref}

\newtheorem{theorem}{Theorem}
\def\wh{\widehat}
\def\wt{\widetilde}
\newcommand{\bbeta}{\bm{\beta}}
\newcommand{\mR}{\mathbb{R}}
\newcommand{\bdelta}{\mbox{\boldmath $\delta$}}

\newcommand{\calL}{\mathcal{L}}
\newcommand{\calD}{\mathcal{D}}

\newcommand{\x}{\mbox{\boldmath $x$}}

\newcommand{\calA}{\mathcal{A}}
\newcommand{\Ah}{\mathcal{A}_h}
 \newcommand{\X}{\mbox{\boldmath $X$}}
 
  \newcommand{\V}{\mbox{\boldmath $V$}}

\newcommand{\w}{\bm{w}}

\newcommand{\trans}{^{\mbox{\tiny{T}}}}
\newcommand{\argmin}{\mathop{\rm arg\min}}
\def\mE{\mathbb{E}}

\newcommand{\mI}{\mathbf{I}}
\newcommand{\mH}{\mathrm{H}}
\newcommand{\mG}{\mathrm{G}}

\begin{document}

\title{Rank-based transfer learning for high-dimensional survival data with application to sepsis data}

\author[1]{Nan Qiao}
\author[1]{Haowei Jiang}
\author[2]{Cunjie Lin} 

\affil[1]{School of Statistics, Renmin University of China, Beijing 100872, China}
\affil[2]{Center for Applied Statistics and School of Statistics, Renmin University of China, Beijing 100872, China \texttt{lincunjie@ruc.edu.cn}}
\date{}
\maketitle

\abstract{Sepsis remains a critical challenge due to its high mortality and complex prognosis. To address data limitations in studying MSSA sepsis, we extend existing transfer learning frameworks to accommodate transformation models for high-dimensional survival data.
Specifically, we construct a measurement index based on C-index for intelligently identifying the helpful source datasets, and the target model performance is improved by leveraging information from the identified source datasets via performing the transfer step and debiasing step. We further provide an algorithm to construct confidence intervals for each coefficient component. Another significant development is that statistical properties are rigorously established, including $\ell_1/\ell_2$-estimation error bounds of the transfer learning algorithm, detection consistency property of the transferable source detection algorithm and  asymptotic theories for the confidence interval construction.  Extensive simulations and analysis of MIMIC-IV sepsis data demonstrate the estimation and prediction accuracy, and practical advantages of our approach, providing significant improvements in survival estimates for MSSA sepsis patients.
}

\keywords{High-dimensional survival data; MIMIC sepsis cohort; Smooth concordance index; U-estimates.}

\label{firstpage}

\section{Introduction}
Sepsis, a life-threatening systemic inflammatory response syndrome caused by infection \citep{evans2021surviving}, affects approximately 50 million people worldwide annually and carries high mortality rates ranging from 15\% to over 50\% \citep{fleischmann2020incidence, rudd2020global}. Due to its high incidence and complex prognosis, sepsis consumes a significant amount of medical resources and incurs substantial expenses. For example, in the USA, sepsis is the most common cause of in-hospital deaths and costs more than \$24 billion annually, accounting for 13\% of healthcare expenditures \citep{paoli2018epidemiology}. In recent years, advancements in sepsis guidelines, timely administration of effective antibiotics, and comprehensive management treatments have led to a decrease in sepsis mortality, but the rate remains alarmingly high \citep{luhr2019trends}.

Sepsis is triggered by the body's extreme response to infection caused by bacteria, fungi, viruses, or parasites.
In sepsis, the most prevalent bacteria that can cause serious clinical consequences are Staphylococcus aureus (S. aureus) and Escherichia coli (E. coli) \citep{Faix:2013}, while S. aureus is a leading cause of bloodstream infections in hospitals. Among these, Methicillin-Susceptible Staphylococcus aureus (MSSA) and methicillin-resistant Staphylococcus aureus (MRSA) have received a lot of attention due to the high prevalence, significant morbidity, and mortality \citep{kourtis2019vital}. Studies also show that Enterococcus and Pseudomonas can induce a severe inflammatory response \citep{marra2006systemic}, while Gram-negative sepsis is also common in clinical study \citep{parker2001experimental}. Besides, other causes of sepsis were also recorded in the MIMIC-IV database (\url{https://mimic.mit.edu}). 
When we only focus on one type of sepsis, say MSSA, the specific data are often inadequate due to the general focus on broader categories of sepsis or antibiotic-resistant strains like MRSA. This may lead to unsatisfactory analysis results, especially when using complex survival models with right-censored data and high-dimensional covariates. In such cases, it is smart to borrow information from other types of sepsis due to their similarities in symptoms, diagnosis, treatment, and prevention. For example, \cite{kavanagh2019control} found that some classical risk factors associated with MRSA infections were also related to MSSA. Also, many studies are conducted to establish differences between clinical, laboratory, and outcomes of MSSA and MRSA infections, which prompts us to be careful when using information from other datasets.

\begin{figure}
    \centering
    \includegraphics[width=0.5\linewidth]{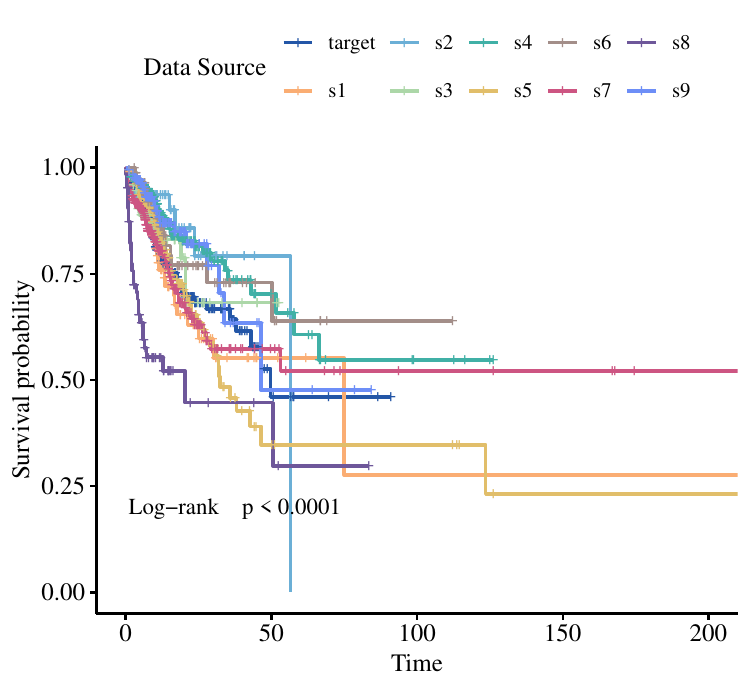}
    \caption{KM curves of different datasets}
    \label{fig:KM-curve}
\end{figure}

Transfer learning, which aims to improve the target task's performance by transferring the knowledge contained in different but related source domains \citep{Radhakrishnan:2023}, is a promising machine learning methodology for solving the above problem. Given the limited data on MSSA sepsis, traditional models based solely on target data often lack accuracy and reliability. By incorporating auxiliary information from sepsis with similar pathophysiological mechanisms, transfer learning is expected to help identify critical features that enhance the understanding and prediction of MSSA sepsis outcomes. Nine different sepsis sources (\Cref{tab:source_desc}) are available for transferring, but we are not quite clear about which ones are useful, since MSSA and other sources share the characteristics of causing severe bacterial infections but differ in bacterial types and antibiotic resistance profiles. A preliminary analysis of the sepsis data (For more details of this dataset, see Section \ref{section-4}) from MIMIC-IV provides strong evidence of both similarities and differences among different types of sepsis (\Cref{fig:KM-curve}). In particular, MSSA is susceptible to methicillin, while MRSA is resistant. Other bacterial sources, such as Streptococcus, Enterococcus, E. coli, and Pseudomonas, vary in Gram stain characteristics, pathogenic mechanisms, treatment protocols, and survival probabilities. Hence, not all sepsis sources are useful for transfer learning due to the differences in resistance profiles and pathogenic mechanisms. This necessitates identifying only helpful datasets to improve parameter estimation and predictive accuracy for MSSA sepsis. 

In recent years, transfer learning approaches have achieved remarkable success in various fields, such as computer vision \citep{wang2018deep}, natural language processing \citep{pruksachatkun2020intermediate},  sentiment analysis \citep{liu2019survey}, image analysis \citep{zhang2015deep}, and the bioinformatics fields \citep{petegrosso2017transfer}. Numerous methodological frameworks have been developed for transfer learning. One key approach is addressing distribution shift \citep{uehara2020off,mo2021learning,wu2023transfer,chu2023targeted} by leveraging summary information, such as moments, from the source population to improve the estimates of the target population. However, traditional distribution shift studies typically need data distribution assumptions between target and source datasets and thus usually require low-dimensional settings, which can limit their practical applicability. Alternatively, several studies have focused on parameter-transfer learning for different statistical models under high-dimensional setting. \cite{li2022transfer} proposed a multi-source transfer learning framework for high-dimensional linear models. \cite{tian2023transfer} extended this framework to generalized linear models and developed a consistent procedure for identifying transferable sources. \cite{qiao2023transfer} conducted transfer learning algorithms for high-dimensional quantile regression (QR) models with the technique of convolution-type smoothing. In semiparametric frameworks,  \cite {hu2023optimal} developed a model averaging approach to transfer parameter information from source models to the target model for prediction. \cite{li2023accommodating} addressed time-varying differences in regression coefficients and baseline hazard functions between a target and a source by developing a transfer learning approach in the Cox proportional hazards model. However, despite these advancements, challenges remain, particularly in survival analysis with high-dimensional predictors and multiple sources, where estimators can become unstable, especially with small sample sizes.

In this study, we propose a transfer learning algorithm to enhance model performance on a target survival dataset by leveraging information from other source datasets with similar but not exactly the same distributions. Instead of Cox and some parametric models, we consider the nonparametric transformation model due to its flexibility for modeling censored survival data \citep{song2007semiparametric}. This model directly exploits the monotonic relationship between survival time and covariates while makes no parametric assumptions on either the transformation function or the error. It includes the Cox and many other models as special cases and is more robust against model misspecification. However, to achieve the transfer learning for high dimensional transformation models, we need to overcome the following difficulties: First, the partial rank (PR) estimation  \citep{khan2007partial}, based on maximizing a discontinuous objective function, is often used to estimate the parameters for a single dataset, which is computationally expensive and practically impossible to compute in the case of high dimensional covariates, not to mention the case with multiple heterogeneous source datasets. Second, we need to intelligently identify the helpful source datasets to avoid the ``negative transfer'' situation where the target task is compromised by the poor source data which is dramatically different from the target cohort. But it is not easy to construct a measurement index for quantifying the usefulness of the source data due to the right censored data and unknown transformation function. Third, although transfer learning has been successfully applied to linear regression model or quantile regression model, the procedure {and its theoretical understanding} in the context of the survival transformation model based on U-estimation is significantly more complicated than the transfer learning based on M-estimation. Accordingly, { new and more challenging theoretical and numerical developments} for identifying informative sources and performing transfer learning with U-estimation are needed. Overall, this study can provide a practically useful new transfer learning approach for survival data with high dimensional covariates when some source data may not improve model performance and can even be harmful. 


The rest of this article is organized as follows. Section \ref{section-2} presents the notation, model, algorithm, { and the corresponding theories are also provided.} Section \ref{section-3} contains the results of our simulation studies, evaluating empirical performance. In Section \ref{section-4}, we provide the data analysis results for the motivating example with MIMIC sepsis cohorts and interpret the findings. Concluding remarks and discussions are provided in Section \ref{section-5}. {All proofs are in Supplementary Materials.}

\section{Methodology}
\label{section-2}

\subsection{Transformation Model}

Let $T$ be an uncensored survival time of interest measured from an initial event to the failure event, which is subject to right censoring. We denote the censoring time by $C$ and the observed survival data are $Y=\min(T,C)$ and $\Delta=I(T\leq C)$. Consider the following transformation model:
\begin{eqnarray}\label{eq:transformation}
    g(T)=\bbeta^\top\X+\varepsilon,
\end{eqnarray}
where $g(\cdot)$ is a monotone increasing function with an unspecified form,  $\varepsilon$ is a random error term with an unknown distribution $F$, which is independent of covariates $\X \in \mathbb{R}^p$, and $\bbeta=(\beta_1,...,\beta_p)^\top$ is a vector of unknown regression parameters with dimension $p$. For identifiability, the $\ell_2$ norm of $\bbeta$ is restricted to 1, that is $\|\bbeta\|_2=(\sum_{j=1}^p |\beta_j|^2)^{\frac{1}{2}}=1$.
Without specifying the form of $g(\cdot)$ and $F$, the transformation model includes many popular models as special cases \citep{chen2002rank}, such as the proportional hazard model, proportional odds model, and accelerated failure time model.

We first briefly review the approach for estimating parameters of model \eqref{eq:transformation} with a single dataset consisting of $n$ independent and identically distributed samples $\calD=\left\{ Y_i,\Delta_i,\X_i\right\}_{i=1}^n$. With unknown $g(\cdot)$ and $F$, the most commonly used approach for estimating the coefficients $\bbeta$ is PR estimation \citep{khan2007partial} with the objective function 
\begin{eqnarray}
   \wh H(\bbeta)=\frac{1}{n(n-1)}\sum_{i \ne j}\Delta_j I(Y_i > Y_j)I(\bbeta^\top\X_i > \bbeta^\top\X_j ),
 \label{eq:trans_loss} 
\end{eqnarray}
where $I(\cdot)$ is the indicator function. Note that $\wh H(\bbeta)$ is a second-order U-statistic quantifying correlation between the observed and fitted values, which differs fundamentally from M-estimation. And the indicator function is non-differentiable, posing more challenges for optimization. To tackle the computational problem, we consider the objective function with a smooth approximation:
\begin{eqnarray*}
	\wh\calL(\bbeta)=\frac{1}{n(n-1)}\sum_{i \ne j}\Delta_j I(Y_i > Y_j) S_n(\bbeta^\top\X_i -\bbeta^\top\X_j ), 
\end{eqnarray*}
where $S_n(x)=\frac{1}{1+\exp(-x/\sigma_n)}$ is used to approximate the indicator function $I(x>0)$, and $\sigma_n$ is strictly positive and decreasing and satisfies $\lim_{n\to\infty} \sigma_n=0$.
By maximizing $\wh\calL(\bbeta)$, we can derive the smoothed partial rank (SPR) estimator, which has been shown to be asymptotically equivalent to the PR estimator \citep{song2007semiparametric}. Thus, the smoothing approach exhibits favorable theoretical properties and yields a computationally affordable estimate with no loss of efﬁciency.

With high dimensional covariates, a popular approach for performing variable selection and regularized estimation is to consider the penalized objective function 
\begin{eqnarray*}
 \wh Q_n(\bbeta)= -\wh\calL(\bbeta)  + p_\lambda(\bbeta),
\end{eqnarray*}
where $p_\lambda(\cdot)$ is the penalty function that depends on the tuning parameter $\lambda$. But it is computationally challenging to optimize $\wh Q_n(\bbeta)$ due to its non-convexity. In this study, we propose using 
the Forward and Backward Stagewise (Fabs) algorithm \citep{shi2018forward}, an effective computational solution to the penalized SPR estimation, to achieve satisfactory computational efficiency and accuracy.

\subsection{Transfer Learning Method}\label{section:transfer}

In the following, we present our method for the transformation model in the framework of transfer learning more formally. Specifically, two types of datasets are observed: a target dataset $\calD^{(0)} = \left\{Y^{(0)}_i,\Delta^{(0)}_i,\X^{(0)}_i \right\}_{i=1}^{n_0}$ and $K$ dependent source datasets with the $k$-th source denoted as $\calD^{(k)} =  \left\{ Y^{(k)}_i,\Delta^{(k)}_i,\X^{(k)}_i\right\}_{i=1}^{n_k}$, $k=1,\cdots, K,$ where $Y_i^{(k)}=\min(T_i^{(k)}, C_i^{(k)})$ and $\Delta_i^{(k)}=I(T_i^{(k)}\leq C_i^{(k)})$, $i=1,\cdots, n_k$. We assume that the transformation models hold in the target and source datasets:
\begin{eqnarray*}
	g(T^{(k)}_i)=\bbeta^{(k) \top}\X^{(k)}_i+\varepsilon_i^{(k)}, ~k=0, 1,\ldots,K, ~ i = 1, \ldots, n_k,
\end{eqnarray*}
where $\bbeta^{(k)} \in \mR^p$ are possibly different coefficients of different cohorts and $g(\cdot)$ is an unspecified monotone increasing function. Here, we assume that both target and source datasets share the same transformation function $g(\cdot)$ for simplicity. On the one hand, in practice, the source domain used for transfer often possesses some similarity with the target domain, so sharing the same transformation function is a reasonable assumption. Under this assumption, model-based transfer learning is reduced to the transfer of parameters, which can simplify the problem. On the other hand, we focus on the estimation of parameters $\bbeta$ and the estimation procedure does not involve the form of transformation function but only requires the function to be monotonic, thus different functions $g_k(\cdot)$ are also allowed in the estimation.

The goal of this study is to transfer useful information from source data to improve the estimation of the target model, which is assumed to be $\ell_0$-sparse in the sense that $\|\bbeta^{(0)}\|_0 = s \ll p$.  Besides, to avoid identifiability problems, we restrict $\|\bbeta^{(k)}\|_2=(\sum_{j=1}^p |\beta_j^{(k)}|^2)^{1/2}=1$. 
When transferring the information from source datasets, we allow the parameters $\bbeta^{(k)}$ to be different from the target parameter $\bbeta^{(0)}$. However, when the target and source models are disparate, the information borrowed from the source datasets may negatively impact the estimation of the target model, a phenomenon known as ``negative transfer". Thus, to ensure effective transfer learning, we need that $\bbeta^{(0)}$ are similar to $\bbeta^{(k)}$, in the sense that $\|\bbeta^{(0)}-\bbeta^{(k)}\|_2\leq h$ for some reasonably small $h>0$. Here,
 $h$ can be treated as a similarity measure. Given a reasonable $h$, the index of helpful datasets is denoted by $\Ah=\{k: k=1,\cdots,K, \|\bbeta^{(0)}-\bbeta^{(k)}\|_2\leq h\}$. In practice, the value of $h$ and the corresponding index $\Ah$ are unknown, and identifying helpful datasets is crucial to ensure transfer gains.

For simplicity, we first consider that the set of helpful datasets $\Ah$ is known and denote $n_{\Ah} = \sum_{k \in \Ah} n_k$, and $n = n_0 + n_{\Ah}$. In general framework of  transfer learning, $n_{\Ah}\gg n_0$. We consider high-dimensional scenarios, i.e., allowing $p > n$.
To estimate $\bbeta^{(0)}$ more effectively and accurately, we propose to borrow information from the datasets $\Ah$ in the following two steps:

{\bf In the first step}, we perform an initial transfer estimation by fitting a transformation model with $\ell_1$-penalty, utilizing data from the target dataset and the helpful source datasets indexed by $\Ah$. That is to compute $\wh{\w}^{\Ah}=\argmin_{\w \in \mR^p} \wh Q_n^{\Ah}(\w)$, where
\begin{eqnarray}
\label{eq:beta_object}
 \wh Q_n^{\Ah}(\w) = -\sum_{k \in \Ah\cup\{0\}}  \frac{\alpha_k}{n_k(n_k-1)} \sum_{i \ne l} \Delta_l^{(k)}I(Y_i^{(k)} > Y_l^{(k)})S_n\left(\w^\top\X_i^{(k)} -\w^\top\X_l^{(k)}\right) + \lambda_{\w} \|\w\|_1,
\end{eqnarray}
in which,  $\alpha_k=n_k/n$, $n=n_0+\sum_{k\in \Ah}n_k$, $\|\w\|_1=\sum_{j=1}^p |\w_j|$, and $\lambda_{\w}$ is a tuning parameter controlling sparsity. 

{\bf In the second step}, we need to run a debiased estimation. As the source data may differ from the target, the estimator obtained in the first step is likely to be biased. Define $\bbeta^{(0)}={\w}^{\Ah}+\bdelta^{\Ah}$, where $\bdelta^{\Ah}$ quantifies the potential difference between the target coefficient $\bbeta^{(0)}$ and fusion parameter $\w^{\Ah}$. In this step, we use only the target data $\calD^{(0)}$ to learn $\bdelta^{\Ah}$, and impose an $\ell_1$-penalty to achieve data-driven debiasing. Specifically, calculate  $\wh{\bdelta}^{\Ah}=\argmin_{\scriptstyle \bm{\delta}  \in \mR^p}\wh Q_{n_0}^{\Ah}(\bdelta)$ with
\begin{eqnarray}\label{eq:delta_object}
	\wh Q_{n_0}^{\Ah}(\bdelta)  = \frac{-1}{n_0(n_0 - 1)} \sum_{i \neq l}\Delta_l I \left(Y^{(0)}_i  > Y^{(0)}_l\right) 
  S_n  \left( (\wh{\w}^{\Ah}  + \bdelta)^\top\X^{(0)}_i 
  -  (\wh{\w}^{\Ah} + \bdelta)^\top  \X^{(0)}_l \right)
  + \lambda_{\scriptstyle \bm{\delta}} \|\bdelta\|_1,
\end{eqnarray}
where $\|\bdelta\|_1=\sum_{j=1}^p\delta_j$, and $\lambda_{\scriptstyle \bm{\delta}}$ is a tuning parameter. Then the final estimator is obtained by $\wh{\bbeta}^{(0)}_{\Ah}=\wh{\w}^{\Ah}+\wh{\bdelta}^{\Ah}$. 

Here we use the Forward and Backward Stagewise (Fabs) algorithm \citep{shi2018forward} to solve \eqref{eq:beta_object} and \eqref{eq:delta_object}, and choose the final solution according to the BIC criterion. {Besides, the SPR estimator introduces a smoothing parameter $\sigma_n$, which will intuitively affect the convergence rate of the final estimate. In numerical study, we take $\sigma_n=n^{-1/2}$ as suggested by \cite{song2007semiparametric} and \cite{shi2018forward}. Specifically, \cite{shi2018forward} rigorously demonstrated that the SPR estimator with the Fabs algorithm is robust to moderate variations in $\sigma_n$ and $\sigma_n=1/\sqrt{n}$ achieves stable performance across high-dimensional settings without requiring case-specific tuning. In fact, we can show that the smoothing bias \citep{tan2022high, qiao2023transfer} is $\mathcal{O}(\sigma_n^2)$ and then the estimation error bounds of the estimator will not be affected by $\sigma_n$ when $\sigma_n = o(n^{-1/4})$. Simulation study in Supplementary Materials also shows that $n^{-1/2}$ is a reasonable choice for $\sigma_n$ in our setting.}

As discussed above, the estimates in \eqref{eq:beta_object} and \eqref{eq:delta_object} are applicable only if the appropriate sources for transfer are known. Thus, we refer to this method as ``Oracle-Trans". In practice, however, transferring information from certain sources may not enhance the target model's performance and can even degrade it. Thus, it is crucial to identify the beneficial datasets. Generally, a source dataset is considered helpful, or transferable, if incorporating its data improves the target model's learning. Here we propose a novel and computationally feasible method to identify informative sources $\wh{\calA}_h$ using concordance index (C-index) \citep{khan2007partial} as the detection criterion. Note that maximizing \eqref{eq:trans_loss} is equivalent to maximizing the C-index function:
\begin{eqnarray} \label{eq:cindex}
  C(\bbeta; \calD) = \frac{\sum_{i \neq j} \Delta_j I(Y_i > Y_j) I(\bbeta^\top \X_i > \bbeta^\top \X_j) }{\sum_{i \neq j}\Delta_j I(Y_i > Y_j) },
\end{eqnarray}
which counts concordant pairs between the predicted and true outcomes and evaluates the overall performance of the fitted survival model. The value of C-index is between 0 and 1, and 1 means an ideal prediction. Thus, we use the C-index to evaluate the performance of the transfer learning.

Next, we propose to detect the informative sources using a three-fold cross-validation approach, which splits the target samples into three parts, performs transfer learning using two parts, and calculates the C-index with the other part:

\underline{Step 1. Split the target data}:  Randomly divide the target dataset $\calD^{(0)}$ into three folds $\calD^{(0)[r]} = \left\{ Y_i^{(0)[r]},\Delta_i^{(0)[r]},\X_i^{(0)[r]}\right\}$ for $r=1,2,3$. 

 \underline{Step 2. Calculate the threshold}: For the $r$-th fold, use $\calD^{(0)[r]}$ as the testing set, and the other two folds, $\calD^{(0)}/\calD^{(0)[r]}$, as the training set. We calculate the estimate $\wh\w^{(0)[r]}$ by minimizing \eqref{eq:beta_object} using the training data $\calD^{(0)}/\calD^{(0)[r]}$ but without using any source dataset, and then calculate the C-index on the testing set $\calD^{(0)[r]}$, i.e., $\wh{C}^{(0)[r]}= C(\wh{\w}^{(0)[r]}; \calD^{(0)[r]})$. Finally, we take the average $\wh{C}^{(0)}= 1/3 \sum_{r=1}^3 \wh{C}^{(0)[r]}$ as the threshold.
 
\underline{Step 3. Calculate the C-index for each source dataset}: We run a transfer estimation using each source dataset to calculate the C-index. For $k= 1,\cdots, K$ and $r=1,2,3$, we obtain a transfer estimate $\wh{\w}^{(k)[r]}$ by minimizing \eqref{eq:beta_object} using $\calD^{(k)} \bigcup \{\calD^{(0)}/\calD^{(0)[r]}\}$. Then, we calculate the C-index on the testing dataset $\calD^{(0)[r]}$, i.e., $\wh{C}^{(k)[r]} = C(\wh{\w}^{(k)[r]}; \calD^{(0)[r]})$, and then take the average of the three-fold cross-validation, $\wh{C}^{(k)}= 1/3 \sum_{r=1}^3 \wh{C}^{(k)[r]}$, $k =1, 2,\ldots,K$ as the C-index for each source data.

\underline{Step 4. Select the informative sources via C-index:} If $\wh{C}^{(k)} > \wh{C}^{(0)}$, we conclude that the $k$-th source dataset is beneficial and include $k$ in $\wh{\calA}_h$. 

{Here, we choose three-fold cross-validation as in \cite{tian2023transfer}. To further explore the effect of different numbers of folds in cross-validation, we conduct an additional simulation study in Supplementary Materials. We find that increasing the number of folds has minimal impact on the estimation accuracy but the computational time increases significantly. 
Therefore, three-fold cross-validation provides a reasonable and efficient trade-off in our setting.} We summarize our method in Algorithm \ref{alg:est_A}. With the selected informative sources, we can implement transfer learning by solving the optimization problems \eqref{eq:beta_object} and \eqref{eq:delta_object} with $\Ah$ replaced by $\wh{\calA}_h$. We denote the resulting estimator as $\wh{\bbeta}^{(0)}_{\text{Auto}}$ and the method is termed ``Auto-Trans" to emphasize its ability to automatically identify the informative source datasets. 

\begin{algorithm}[h]
	\caption{Detection of informative sources}  
	\label{alg:est_A}  
	\begin{algorithmic}  
		\Require  
		Target dataset $\calD^{(0)}$ and $K$ source datasets $\calD^{(k)}, k=1,2,\cdots,K$.
		\Ensure $\wh{\calA}_h$, the estimate of $\Ah$.
    	\State Divide the target dataset $\calD^{(0)}$ randomly into three folds, i.e., $\calD^{(0)} =\bigcup_{r=1}^3 \calD^{(0)[r]}$.		
		\For{$r = 1, 2, 3$}
		 \State Solve \eqref{eq:beta_object} using $\calD^{(0)}/\calD^{(0)[r]}$ to obtain $\wh{\w}^{(0)[r]}$;
          \State Calculate the C-index \eqref{eq:cindex} on $\calD^{(0)[r]}$: $\wh C^{(0)[r]}= C(\wh{\w}^{(0)[r]}; \calD^{(0)[r]})$; 
         \State Calculate the threshold $\wh{C}^{(0)}= 1/3 \sum_{r=1}^3 \wh{C}^{(0)[r]}$;

		\For{$k=1, \ldots, K$} 		
		\State Solve \eqref{eq:beta_object} using $\calD^{(k)} \bigcup \{\calD^{(0)}/\calD^{(0)[r]} \}$ to obtain $\wh{\w}^{(k)[r]}$; 
        \State Calculate the C-index \eqref{eq:cindex} on $\calD^{(0)[r]}$: $\wh{C}^{(k)[r]}= C(\wh\w^{(k)[r]}; \calD^{(0)[r]})$.

		\EndFor
		\EndFor
	    \State Initialise $\wh{\calA}_h=\varnothing$;
	    \For{$k=1,...,K$,} 
         \State Calculate $\wh{C}^{(k)}= 1/3 \sum_{r=1}^3 \wh{C}^{(k)[r]}$. 
	    \If {$\wh{C}^{(k)} > \wh{C}^{(0)}$,}
	    \State $\wh{\calA}_h=\wh{\calA}_h \cup\{k\}$ ; 
	    \EndIf
	    \EndFor		    	
	\end{algorithmic}  
\end{algorithm}

{
\subsection{Theoretical Properties and Confidence Interval Construction}
\subsubsection{Estimation error bound and detection consistency}
In this section, we establish the theoretical properties of the proposed methods. We first introduce some notations to be used in this text. 
For a vector $\mathbf{v}=(v_1,\cdots,v_p)^\top \in \mR^p$ and $q \in[1, \infty)$, let $\|\mathbf{v}\|_q=\left(\sum_{j=1}^p\left|v_j\right|^q\right)^{\frac{1}{q}}$ be its $\ell_q$ norm, $\|\mathbf{v}\|_0 = \# \{j: v_j \neq 0\}$ be its $\ell_0$ norm and $\|\mathbf{v}\|_{\infty} = \max_{1\leq j \leq p}|v_j|$ be its $\ell_{\infty}$ norm. For a matrix $\mathbf{A}_{p \times q} =  [a_{ij}]_{p \times q}$, 
let 
$\|\mathbf{A}\|_\infty = \max_{1\leq i \leq p} \sum_{j=1}^{q} |a_{ij}|$, 
and $\|\mathbf{A}\|_{\max} = \max_{i,j} |a_{ij}|$. Let $\lambda_{\min}(\mathbf{A})$ and $\lambda_{\max }(\mathbf{A})$ be its eigenvalue with the smallest value and largest value, respectively. This is the common notation for eigenvalues of a matrix, and $\lambda_{\min}, \lambda_{\max}$ should not be confused with the penalization parameter used in a penalty function.
For a sequence $\big\{a_n\big\}$ and another nonnegative sequence $\big\{b_n\big\}$, we write $a_n=\mathcal{O}(b_n)$ or $a_n \lesssim b_n$ if there exists a constant $c>0$ such that $\left|a_n\right| \leq c b_n$ for all $n \geq 1$. 
 Also, we use $a_n=o\left(b_n\right)$ or $a_n \ll b_n$ to represent $\lim_{n \rightarrow \infty} \frac{a_n}{b_n}=0$. We write $b_n \gg a_n$ if $a_n \ll b_n$. All proofs are in Supplementary Materials.

The following Theorem \ref{th1} gives the $\ell_1 / \ell_2$-estimation error bound for the Oracle-Trans estimator, which is based on the known informative sources.

\begin{theorem}[$\ell_1 / \ell_2$-estimation error bound of Oracle-Trans] \label{th1}
Assume Conditions (C1) (C2) (C3)
in Supplementary Materials hold, $n_0 > Cs^2\log p$, and $h \leq s\sqrt{\log p/n_0}$, where $C > 0$ is a constant. 
We take $\lambda_{\w}=C_{\w} \sqrt{\frac{\log p}{n_{\Ah}+n_0}}$ and $\lambda_{\scriptstyle \bm{\delta}}=C_{\scriptstyle \bm{\delta}} \sqrt{\frac{\log p}{n_0}}$, where $C_{\w}$ and $C_{\scriptstyle \bm{\delta}}$ are sufficiently large positive constants, then
\begin{align}
 \|\wh{\bbeta}^{(0)}_{\Ah} - {\bbeta}^{(0)}\|_2 
& \lesssim  h^{1/2} \left(\frac{\log p}{n_0}\right)^{1/4} + s^{1/2} \left(\frac{\log p}{n_0}\right)^{1/4}\left(\frac{\log p}{n_0 + n_{\Ah}}\right)^{1/4},\label{eq:l2bound}\\
  \|\wh{\bbeta}^{(0)}_{\Ah} - {\bbeta}^{(0)} \|_1
&\lesssim s \left(\frac{\log p}{n_{\Ah}+n_0}\right)^{1/2} + \left(\frac{\log p}{n_0 + n_{\Ah}}\right)^{1/4} (sh)^{1/2} + h,
\label{eq:l1bound}
\end{align}
with probability at least $1 - 2p^{-1}$. 
\end{theorem}

Theorem \ref{th1} implies that, when $\Ah$ is an empty set, the upper bound in \eqref{eq:l2bound} is $\mathcal{O}_P (\sqrt{s\log p/n_0})$. When $\Ah$ is non-empty, the upper bound in \eqref{eq:l2bound} is sharper than $\sqrt{s\log p/n_0}$ and the upper bound in \eqref{eq:l1bound} is sharper than $s\sqrt{\log p/n_0}$ if $ n_0 \lesssim n_{\Ah}$ and $h < s\sqrt{\log p/n_0}$. Similar to Theorem 4 of \cite{tian2023transfer}, we can show the following detection consistency for Algorithm \ref{alg:est_A}.

\begin{theorem}[Detection consistency of $\Ah$]\label{th2}
   For Algorithm \ref{alg:est_A}, with Condition (C4) in Supplementary Materials satisfied for some $h$, for any $\delta> 0$, there exist constants $C^\prime(\delta)$ and $N = N(\delta) > 0$ such that when $M_1 = C^\prime(\delta)$ and $\min_{k \in \{0\} \cup \Ah } n_k > N(\delta)$,we have $
  \mathcal{P}(\wh{\calA}_h = \Ah) \geq 1 - \delta.$
\end{theorem}

\subsubsection{Confidence interval construction}

 In this section, we construct the asymptotic confidence interval (CI) for each component of $\bbeta^{(0)}$ based on the previous transfer learning estimator $\wh{\bbeta}^{(0)}_{\Ah}$.
Motivated by \cite{cai2024statistical}, \cite{zhang2014confidence} and \cite{ning2017general}, we consider the desparsified estimator 
$\wh{\bbeta}^{(0)}_{\Ah} + \mH^{-1} \wh{\bm{\eta}},$
where ${\mH} = - \mE \Big\{ \Delta_l^{(0)} I(Y_i^{(0)} > Y_l^{(0)}) S_n^{\prime\prime}\Big(\bbeta^{(0)}(\X_i^{(0)} - \X_l^{(0)})\Big) (\X_i^{(0)} - \X_l^{(0)}) (\X_i^{(0)} - \X_l^{(0)})^\top \Big\}$ is the inverse Hessian matrix, and $\wh{\bm{\eta}} :=  \frac{1}{n_0(n_0-1)} \sum_{i \ne l}  \Delta_l^{(0)} I(Y_i^{(0)} > Y_l^{(0)}) S_n^{\prime}\left(\wh{\bbeta}^{(0)}_{\Ah}(\X_i^{(0)} - \X_l^{(0)})\right) (\X_i^{(0)} - \X_l^{(0)})$ is the negative gradient. 
Unfortunately, $\mH$ is unknown in the above formula. Even if we can estimate $\mH$ by $\wh\mH$, we cannot estimate $\mH^{-1}$ directly by $\wh\mH^{-1}$, since the matrix $\wh\mH$ may not be invertible when the dimension $p$ is larger than the sample size $n_0$. To address this, we adopt the approach in \cite{cai2011constrained} to obtain $\wh\mH^{-1}$. Denote the estimator of $\mH^{-1}$ as $\wh{\Theta}$. Then we obtain $\wh{\Theta}$ via the following convex program:
\begin{eqnarray}\label{eq:Theta_solution}
\min_{\Theta \in \mathbb{R}^{p \times p}} \|\Theta\|_{\infty}, \quad \text{s.t.} \quad \|\Theta \wh\mH - \mI \|_{\max} \leq \gamma_n.
\end{eqnarray}
Following \cite{cai2011constrained}, we use five-fold cross-validation to select $\gamma_n$. Finally, the desparsified estimator is defined as follows:
\begin{eqnarray}\label{eq:debiased_solution}
\wt{\bbeta} = \wh{\bbeta}^{(0)}_{\Ah} + \wh{\Theta} \wh{\bm{\eta}}.
\end{eqnarray}
The details about confidence interval construction are shown in Algorithm 2, and the corresponding theories are provided in Theorem \ref{th3}.

\begin{algorithm}[h]\label{algorithm:CI}
\caption{Confidence interval construction for the high-dimensional transformation model}
    \begin{algorithmic}[1]%
\Require{Target dataset $\calD^{(0)} = \left\{Y^{(0)}_i,\Delta^{(0)}_i,\X^{(0)}_i \right\}_{i=1}^{n_0}$; transferring estimator $\wh{\bbeta}^{(0)}_{\Ah}$;}
\Ensure{Desparsified Lasso estimator $\wt{\bbeta}$ and its confidence intervals  
$\{\mathcal{I}_j\}_{j=1}^p$  } ; 

\item Compute the negative gradient $\wh{\bm{\eta}}$ and Hessian matrix $\wh{\mH}$ using the target data:
\begin{eqnarray*}
    \wh{\bm{\eta}} =  \frac{1}{n_0(n_0-1)} \sum_{i \ne l} \Delta_l^{(0)} I\left(Y_i^{(0)} > Y_l^{(0)}\right) S_n^{\prime}\left(\wh{\bbeta}^{(0)\top}_{\Ah}(\X_i^{(0)} - \X_l^{(0)})\right) (\X_i^{(0)} - \X_l^{(0)}),
\end{eqnarray*}
and
\begin{eqnarray*}
    \wh{\mH} =  -\frac{1}{n_0(n_0-1)} \sum_{i \ne l}  \Delta_l^{(0)} I\left(Y_i^{(0)} > Y_l^{(0)}\right) S_n^{\prime\prime}\left(\wh{\bbeta}^{(0)\top}_{\Ah}(\X_i^{(0)} - \X_l^{(0)})\right) (\X_i^{(0)} - \X_l^{(0)}) (\X_i^{(0)} - \X_l^{(0)})^\top,
\end{eqnarray*}
where $S^{\prime}_n(x) = S_n(x)(1-S_n(x))/\sigma_n$ and $S^{\prime\prime}_n(x) = S_n(x)(1-S_n(x))(1-2S_n(x))/\sigma_n^2$. 

\item Compute $\wh{\Theta}$ by solving the optimization problem (\ref{eq:Theta_solution}).

\item Compute the desparsified estimator:
$\wt\bbeta = \wh{\bbeta}^{(0)}_{\Ah} +  \wh{\Theta}  \wh{\bm{\eta}}.$

    \item Construct the confidence interval for $\beta_j^{(0)}$, $j = 1, \ldots,p$:
    \begin{eqnarray*}
    \mathcal{I}_j \gets \left[\wt{\beta}_j - \sqrt{\wh\Theta_j^\top \wh{\mG} \wh\Theta_j} q_{\alpha/2}/\sqrt{n_0}, \, 
    \wt{\beta}_j + 
    \sqrt{\wh\Theta_j^\top \wh{\mG} \wh\Theta_j} q_{\alpha/2}/\sqrt{n_0}\right],
    \end{eqnarray*}
 where $
\wh{\mG} = n_0^{-1} \sum_{l = 1}^{n_0} \{\nabla \hat{\tau}_n(\V_l^{(0)}, \wh{\bbeta}^{(0)}_{\Ah}) \nabla \hat{\tau}_n^\top(\V_l^{(0)}, \wh{\bbeta}^{(0)}_{\Ah})\}
$, with $\nabla \hat{\tau}_n(v, \bbeta) = n_0^{-1} \sum_{i = 1}^{n_0} \{\Delta^{(0)}_i I(y \geq Y^{(0)}_i)S^\prime_n(\bbeta^\top \x - \bbeta^\top \X^{(0)}_i)(\x - \X^{(0)}_i)^\top+\delta I(Y^{(0)}_i \geq y)S^\prime_n(\bbeta^\top \X^{(0)}_i - \bbeta^\top \x)(\X^{(0)}_i-\x )^\top\}$, in which $\V_i^{(0)} = (\Delta_i^{(0)}, Y^{(0)}_i, \X^{(0)}_i)$ and $v = (\delta, y, x)$. 
Here $\wt{\beta}_j$ is the $j$-th component of $\wt{\bbeta}$, and $q_{\alpha/2}$ is the $\alpha/2$-left tail quantile of $\mathcal{N}(0, 1)$.
    \item Output the confidence intervals $\{\mathcal{I}_j\}_{j=1}^p$.
\end{algorithmic}
\end{algorithm}

\begin{theorem} \label{th3}
Suppose that $\wh{\bbeta}^{(0)}_{\Ah}$ satisfies the estimation error bound in Theorem 1 with $h \leq s\sqrt{\log p/n_0}$, based on Conditions (C1) and (C5), as $n_0\to \infty$, we have 
\begin{eqnarray}\label{eq:normal}
    \sqrt{n_0}(\wt\bbeta - {\bbeta}^{(0)}) \xrightarrow{d} \mathcal{N}(0 ,\mH^{-1\top }{\mG} \mH^{-1}),
\end{eqnarray}
where $\mG$ is the asymptotic variance of $\wh{\bm{\eta}}({\bbeta}^{(0)})$, ${\mG} = \mE \{\nabla {\tau}(\V_i^{(0)}, {\bbeta}^{(0)}) \nabla {\tau}^\top(\V_i^{(0)}, {\bbeta}^{(0)})\}
$ with ${\nabla\tau}(v, \bbeta) =\mE\left\{ \Delta^{(0)}_i I(y \geq Y^{(0)}_i)S_n'(\bbeta^\top \x -\bbeta^\top \X^{(0)}_i)(x-\X_i^{(0)})+\delta I(Y^{(0)}_i \geq y)S_n'(\bbeta^\top \X^{(0)}_i -\bbeta^\top \x)(\X_i^{(0)}-x)\right\}$, in which $\V_i^{(0)} = (\Delta_i^{(0)}, Y^{(0)}_i, \X^{(0)}_i)$ and $v = (\delta, y, x)$. 
\end{theorem} 

In the proof of Theorem 3, we show that the desparsified estimator $\wt\bbeta$ enjoys the following Bahadur representation:
    \begin{align*}
   \| \sqrt{n_0}(\wt\bbeta - {\bbeta}^{(0)}) - \sqrt{n_0} \mH^{-1} \wh{\bm{\eta}}({\bbeta}^{(0)}) \|_{\infty} =\mathcal{O}\Bigg(\sqrt{\log p} \Big(\sqrt{\frac{\log p}{n_0}} \| \wh{\bbeta}^{(0)}_{\Ah}-  \bbeta^{(0)}\|_1 \Big)^{1-q} + \sqrt{\log p}\bigl\|\wh{\bbeta}^{(0)}_{\Ah}- \bbeta^{(0)}\bigr\|_{1}^2\Bigg),%
\end{align*}
where $\|\wh{\bbeta}^{(0)}_{\Ah} - {\bbeta}^{(0)}\|_1 \lesssim s \sqrt{\frac{\log p}{n_{\Ah} + n_0}} + \left(\frac{\log p}{n_{\Ah} + n_0}\right)^{1/4} \sqrt{sh}  + h$ shown in Theorem 1. This suggests that $\sqrt{n_0}(\wt\bbeta - {\bbeta}^{(0)})$ can be expressed as a high-dimensional U-statistic up to some negligible terms. With this help, the asymptotic distribution of the estimators can be derived using the central limit theorem for U-statistic $\wh{\bm{\eta}}({\bbeta}^{(0)})$ \citep{song2007semiparametric,lee2019u,lin2013smoothed}, as shown in Theorem \ref{th3}.

}

\section{Simulation studies}
\label{section-3}

In this section, we conduct comprehensive simulations, evaluate the performance of the proposed approaches, and compare them against multiple alternatives. For simplicity, we refer to the first step of our Oracle-Trans algorithm as the fusion learning step, and the second step as the debiasing step. We compare the following six methods:
\begin{itemize}
    \item {\bf Target-Only:} Perform the naive estimation using only the target dataset.
    \item {\bf Naive-Pooled:} Perform the fusion learning step by pooling all data together.
    \item {\bf Oracle-Pooled:} Assume $\Ah$ is known and performs the fusion learning step by minimizing \eqref{eq:beta_object} using both sources in $\Ah$ and the target dataset without debiasing.
    \item {\bf Oracle-Trans:} Perform the debiasing step using the target dataset after Oracle-Pooled.
    \item {\bf Auto-Pooled:} Run Algorithm \ref{alg:est_A} to obtain $\wh{\calA}_h$ and then performs the fusion learning step by minimizing \eqref{eq:beta_object} using sources in $\wh{\calA}_h$ and the target dataset without debiasing.
    \item {\bf Auto-Trans:} Perform the debiasing step using the target dataset after Auto-Pooled.
\end{itemize}

\subsection{Data generation}

When generating data, we consider several scenarios with varying discrepancies between the target and source models, different numbers of source datasets, and proportions of informative source datasets. We also consider different dimensions and sample sizes. Specifically, we generate target and source datasets from the accelerated failure time (AFT) models:
\begin{eqnarray*}
\log(T_i)=\bbeta^{(k)^\top}\X^{(k)}_i+\varepsilon_i,~ k =0, 1, \ldots, K,~i = 1,\ldots, n_k,
\end{eqnarray*}
where $n_0$ and $n_k$ are sample sizes of the target dataset and the $k$-th source dataset, respectively. For the target dataset, $\X^{(0)}_i \overset{i.i.d.}{\sim} \mathcal{N}(\mathbf{0}_p,\mathbf \Sigma^{(0)})$, $i=1,\cdots, n_0$ and $\mathbf\Sigma^{(0)}=[0.3^{|j-j^\prime|}]$ for $j, j^\prime = 1,\ldots,p$.
 For the $k$-th source dataset, $k=1,\cdots, K$, $\X_i^{(k)} \overset{i.i.d.}{\sim} \mathcal{N}(\mathbf{0}_p,\mathbf \Sigma^{(k)})$, $i=1,\cdots, n_k$, where $\mathbf{\Sigma}^{(k)}=\mathbf\Sigma^{(0)}+\pmb{v}\cdot\pmb v{^\top}$ and $\pmb{v}\sim \mathcal{N}(\mathbf{0}_p, 0.3^2 \cdot \mathrm{I}_p)$, $\mathrm{I}_p$ is a $p$-dimensional identity matrix. The random error $\varepsilon_i \sim \mathcal{N}(0,0.2)$ is independent of $\X_i^{(k)}$. The censoring time is generated from $\mathrm{Exp}(1/\theta)$, where $\theta$ is set to achieve the censoring rate about 40\%. 
 
We consider different dimensions of covariates. In Scenarios S1-S5 and S7, we set $p=200$ and $s=\|\bbeta^{(0)}\|_0=12$ with $\bbeta^{(0)}=(1\cdot\mathbf{1}_2^{\trans},-1\cdot\mathbf{1}_2^{\trans},0.8\cdot\mathbf{1}_2^{\trans},-0.8\cdot\mathbf{1}_2^{\trans}, 0.6\cdot\mathbf{1}_2^{\trans},-0.6\cdot\mathbf{1}_2^{\trans},\mathbf{0}_{p-s}^{\trans})^{\trans}$, where $\mathbf{1}_2$ is a two-dimensional vector of all 1,  $\mathbf{0}_{p-s}$ is a $p-s$-dimensional vector of all 0. 
In Scenario 6, we set $p=500$ and $s=24$ with $\bbeta^{(0)}=(1\cdot\mathbf{1}_4^{\trans},-1\cdot\mathbf{1}_4^{\trans},0.8\cdot\mathbf{1}_4^{\trans},-0.8\cdot\mathbf{1}_4^{\trans}, 0.6\cdot\mathbf{1}_4^{\trans},-0.6\cdot\mathbf{1}_4^{\trans},\mathbf{0}_{p-s}^{\trans})$. In these scenarios, 
the coefficients $\bbeta^{(k)}$ for the source data are generated by perturbing $\bbeta^{(0)}$. 
Specifically, let $\mathcal{J}=\{j:\beta^{(0)}_j\neq 0\}$ and $\mathcal{J}^c=\{j:\beta^{(0)}_j= 0\}$.
For the $k$-th source, we construct three index subsets $\mathcal{J}_1^{(k)}$, $\mathcal{J}_2^{(k)}$ and $\mathcal{J}_3^{(k)}$ with sizes $d_1$, $d_2$ and $r$, respectively. The elements in $\mathcal{J}_1^{(k)}$ and $\mathcal{J}_3^{(k)}$ are randomly selected from $\mathcal{J}$ while the elements in $\mathcal{J}_2^{(k)}$ are randomly selected from $\mathcal{J}^c$. The perturbations are performed on the corresponding coefficients as follows: $\beta^{(k)}_j=\beta^{(0)}_j+\epsilon^{(k)}_j$ for $j\in \mathcal{J}_1^{(k)}$ and $\mathcal{J}_2^{(k)}$;  $\beta^{(k)}_j=-\beta^{(0)}_j$ for $j\in \mathcal{J}_3^{(k)}$, where $\epsilon^{(k)}_j$ are generated from the uniform distribution $U[-u,u]$. To avoid non-identifiability, all coefficients are normalized such that $\|\bbeta^{(k)}\|_2 = 1$, for $k = 0,1,\ldots, K$. For each scenario, $d_1$, $d_2$, $r$ and $u$ are carefully designed to generate informative and non-informative sources. Here, we introduce the estimated rank correlation (ERC) to assess the similarity between the target and source coefficients:
\begin{eqnarray*}
    \mathrm{ERC}(\bbeta^{(0)},\bbeta^{(k)})=\frac{\sum_{i\ne j} I(\beta^{(0)}_i>\beta^{(0)}_j)I(\beta^{(k)}_i>
    \beta^{(k)}_j)}{\sum_{i \ne j}I(\beta^{(0)}_i>\beta^{(0)}_j)},
\end{eqnarray*}
where a larger ERC indicates a more useful dataset. For helpful datasets $k \in {\Ah}$, we set $d_1=2$ or $4$, $d_2 = 4$, $u=0.3$ or $0.4$, $r = 2$, to achieve that the ERC$>0.8$. For unhelpful datasets $k \in {\Ah^c}$, we set $d_1 = 6$, $d_2 = 6$, $u = 1$, $r = 7$, and the corresponding ERC$<0.5$. Specifically, we consider the following seven scenarios:
\begin{itemize}
    \item S1:  $n_0=100$, $n_k=200$, $p=200$, $K=2$, $|\Ah|=1$. For the helpful source, we set $d_1=2$, $d_2=4$, $r=2$, $u=0.3$. For the unhelpful source, we set $d_1=6$, $d_2=6$, $r=7$, $u=1$.
    \item S2:  $n_0=100$, $n_k=200$, $p=200$, $K=2$, $|\Ah|=1$. For the helpful source, we set $d_1=4$, $d_2=4$, $r=2$, $u=0.4$. For the unhelpful source, we set $d_1=6$, $d_2=6$, $r=7$, $u=1$.
    \item S3: $n_0=100$, $n_k=100$, $p=200$, $K=6$, $|\Ah|=3$. Other settings are the same as S1.
    \item S4: $n_0=100$, $n_k=60$, $p=200$, $K=6$, $|\Ah|=3$. Other settings are the same as S1.
   \item S5: $n_0=100$, $n_k=100$, $p=200$, $K=6$, $|\Ah|=2$. Other settings are the same as S1.
    \item S6:  $n_0=200$, $n_k=200$, $p=500$, $K=6$, $|\Ah|=3$. For the helpful source, we set $d_1=4$, $d_2=10$, $r=2$, $u=0.4$. For the unhelpful source, we set $d_1=12$, $d_2=10$, $r=14$, $u=1$.
    \item S7: $n_0=60$, $n_k=60$, $p=200$, $K=10$, $|\Ah|=3,6,9$. For the helpful source, we set $d_1=2$, $d_2=3$, $r=2$, $u=0.3$. For the unhelpful source, we set $d_1=7$, $d_2=7$, $r=9$, $u=1$.
\end{itemize}

In S1, there are two sources with $\mathrm{ERC}(\bbeta^{(0)},\bbeta^{(k)})=0.834$ for $k \in \Ah$ and $0.418$ for $k \in \Ah^c$. Compared to S1, more perturbations are added in S2, while S3 considers more sources and retains the same perturbations as in S2. Compared with S3, the sample sizes of the source datasets in S4 are reduced while the proportion of helpful sources in S5 is reduced. In S6, we consider a setting with higher dimensions $p=500$. Besides, we aim to investigate how the various methods perform when the proportion of helpful source datasets is increased through S7. To evaluate the methods, we consider the following measurements: (1) F1-score for assessing variable selection accuracy; (2) { RMSE of the estimates for assessing estimation accuracy: $\text{RMSE}=\left\{\sum_{j=1}^p(\wh{\beta}_j-\beta_j)^2\right\}^{1/2}$};
(3) C-index $C(\wh{\bbeta};\calD^{*})$ in \eqref{eq:cindex} for evaluating prediction accuracy. It should be noted that the C-index is calculated based on a testing set $\calD^{*}$ of the target domain data, which is generated independently with sample size $n_{*}=30$. {For each scenario, the results are summarized based on 500 replications.}

{Additionally, to evaluate the accuracy of the proposed method for constructing confidence intervals, we also conduct a simulation. We use three metrics for evaluation: (1) the bias of the estimator before and after the debiasing process; (2) the coverage probability of the confidence interval, which is the proportion of times the confidence interval covers the true value in 500 repetitions; (3) the length of the confidence interval. We considered two types of variables: signal variables with nonzero coefficients and noise variables with zero coefficients. Due to space limitations, the details of the scenario and simulation results are placed in Supplementary Materials.}

\subsection{Simulation results}

\begin{figure}[htbp]
    \centering
    \begin{subfigure}[b]{0.3\linewidth}
        \centering
        \includegraphics[width=\linewidth]{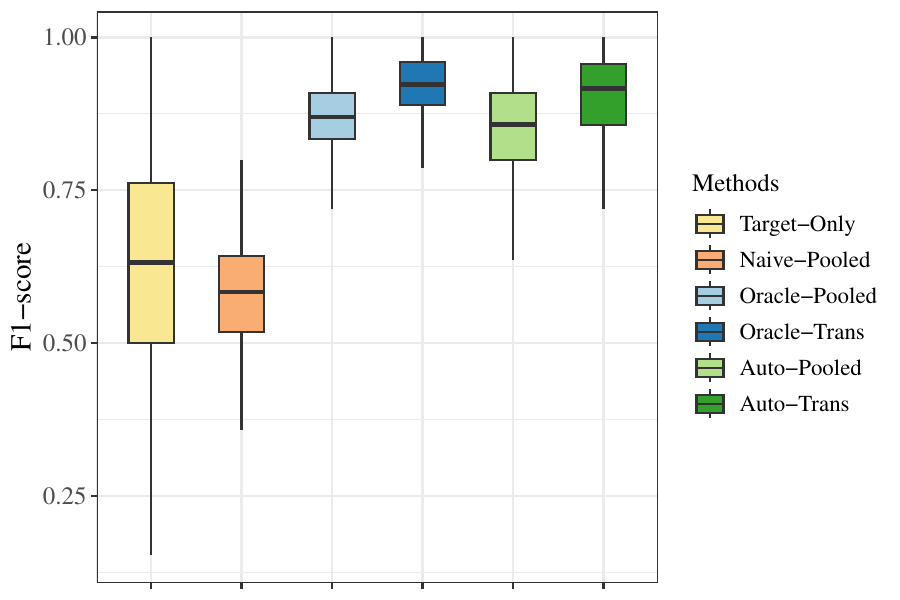}
         \caption{S1}
    \end{subfigure}
    \hfill
    \begin{subfigure}[b]{0.3\linewidth}
        \centering
        \includegraphics[width=\linewidth]{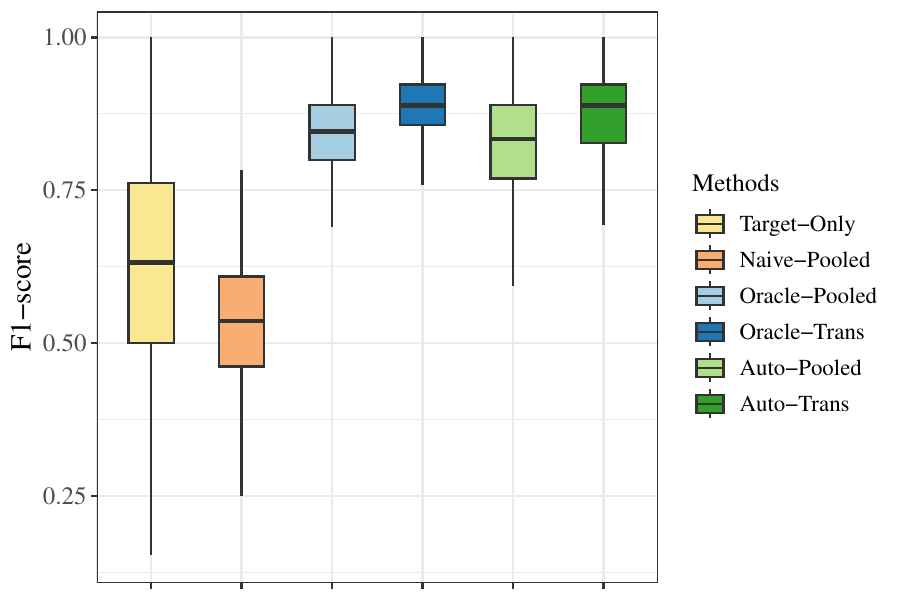}
         \caption{S2}
    \end{subfigure}
    \hfill
    \begin{subfigure}[b]{0.3\linewidth}
        \centering
        \includegraphics[width=\linewidth]{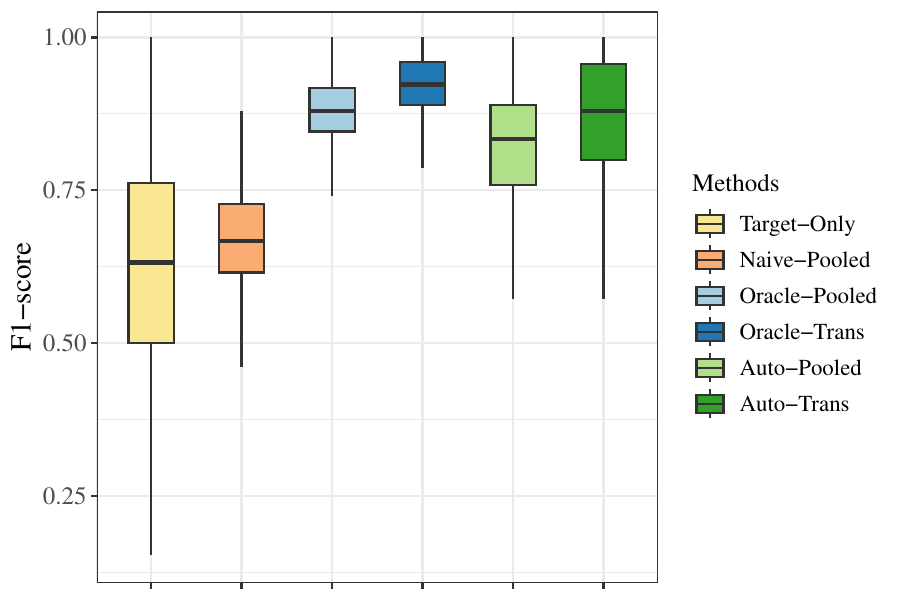}
        \caption{S3}
    \end{subfigure}

     \vspace{\floatsep} 
     
    \begin{subfigure}[b]{0.3\linewidth}
        \centering
        \includegraphics[width=\linewidth]{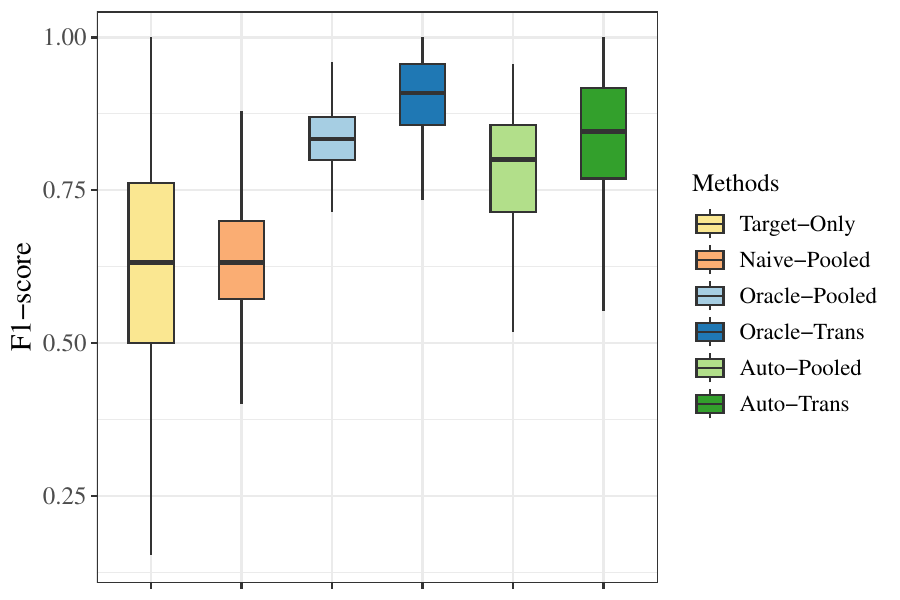}
         \caption{S4}
    \end{subfigure}
    \hfill
    \begin{subfigure}[b]{0.3\linewidth}
        \centering
        \includegraphics[width=\linewidth]{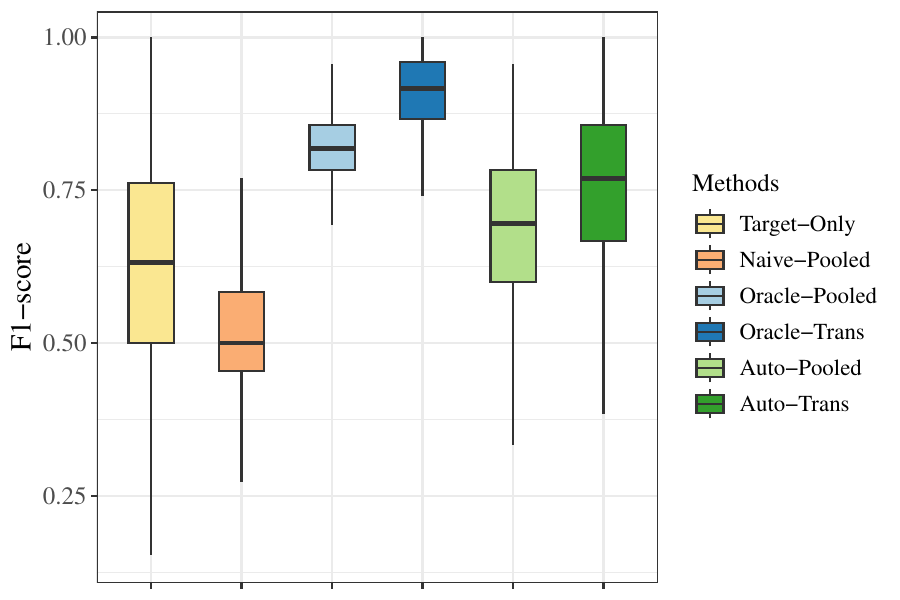}
         \caption{S5}
    \end{subfigure}
    \hfill
    \begin{subfigure}[b]{0.3\linewidth}
        \centering
        \includegraphics[width=\linewidth]{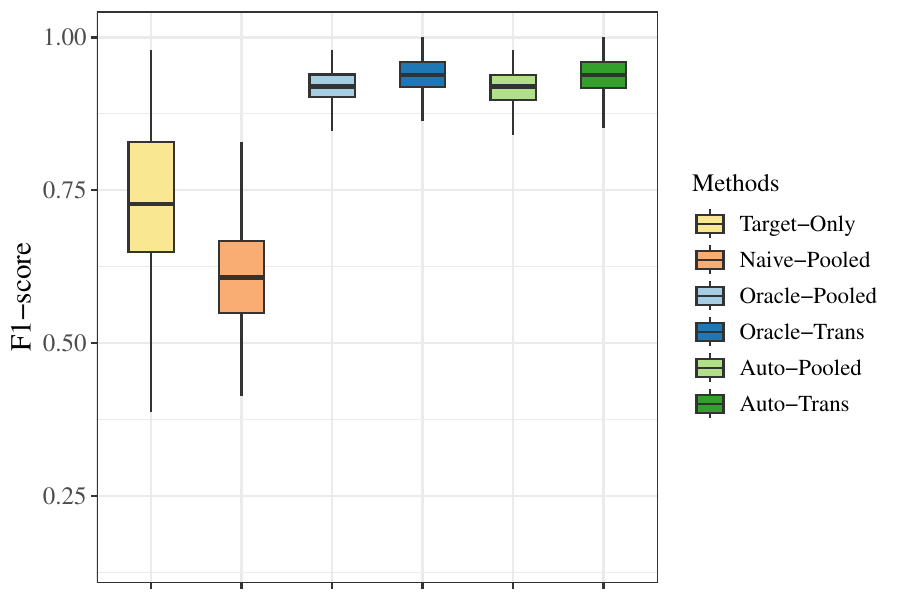}
         \caption{S6}
    \end{subfigure}
    \caption{Simulation results for F1-score under Scenarios S1-S6}
    \label{fig:F1_SCORE}
\end{figure}

Across the entire spectrum of simulation, the proposed Auto-Trans method is observed to have performance either at or near the best in variable selection, estimation accuracy, and prediction accuracy. Specifically, the following conclusions are obtained:

(1). The box-plots of F1-score in Figure \ref{fig:F1_SCORE} show that transfer learning offers advantages in variable selection over the Target-only and Naive-Pooled methods. The four transfer learning related methods produce much larger F1-scores than the other two methods. In almost all settings, Auto-Pooled performs similarly to Oracle-Pooled, while Auto-Trans yields results very close to Oracle-Trans in terms of F1-score. This further reconfirms the accuracy of the detection algorithm for identifying informative sources.

\begin{figure}[htbp]
    \centering
    \begin{subfigure}[b]{0.3\linewidth}
        \centering
        \includegraphics[width=\linewidth]{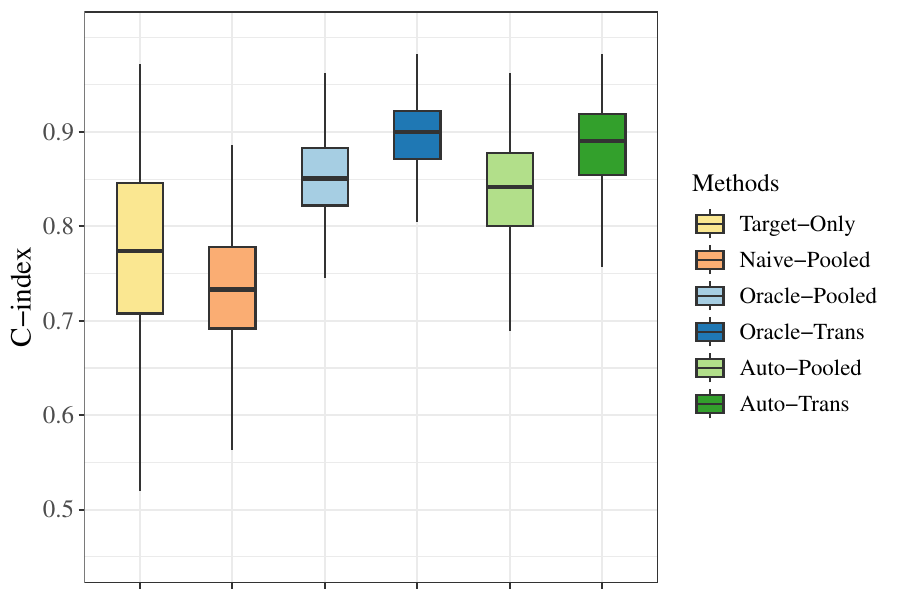}
         \caption{S1}
    \end{subfigure}
    \hfill
    \begin{subfigure}[b]{0.3\linewidth}
        \centering
        \includegraphics[width=\linewidth]{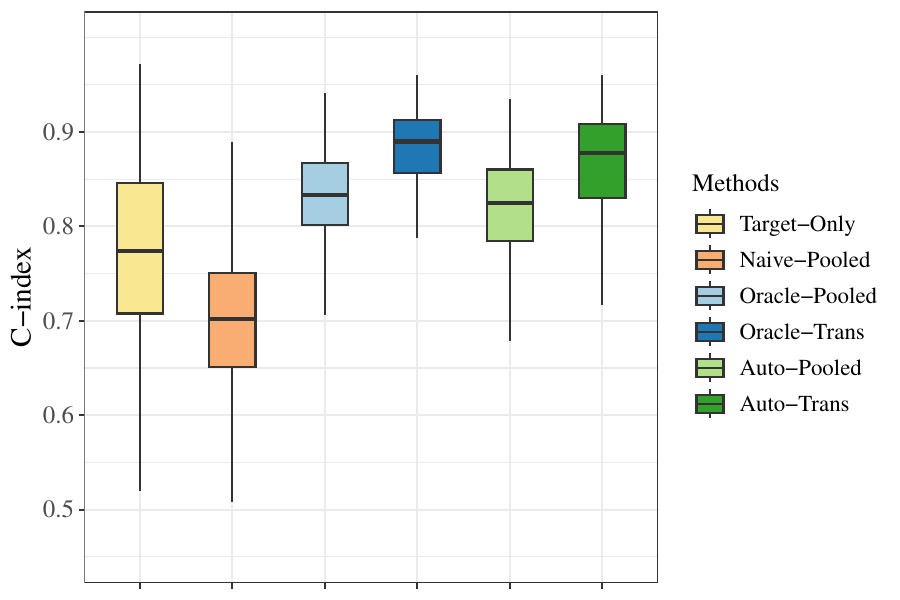}
         \caption{S2}
    \end{subfigure}
    \hfill
    \begin{subfigure}[b]{0.3\linewidth}
        \centering
        \includegraphics[width=\linewidth]{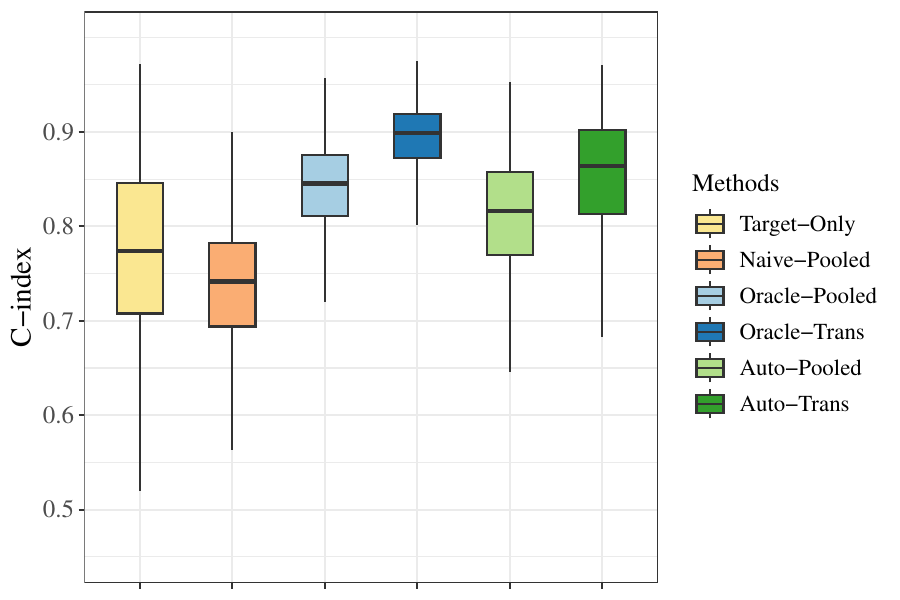}
        \caption{S3}
    \end{subfigure}

     \vspace{\floatsep} 
     
    \begin{subfigure}[b]{0.3\linewidth}
        \centering
        \includegraphics[width=\linewidth]{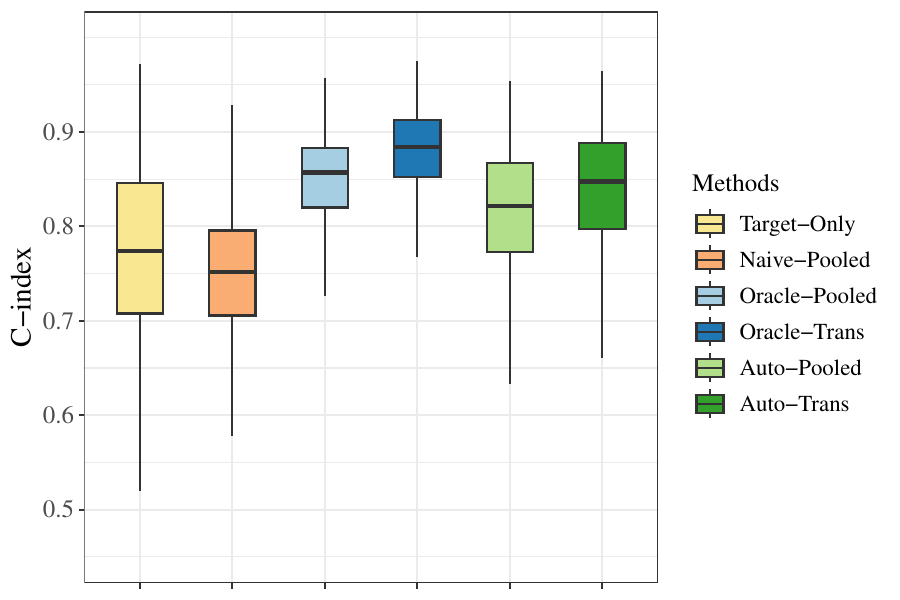}
         \caption{S4}
    \end{subfigure}
    \hfill
    \begin{subfigure}[b]{0.3\linewidth}
        \centering
        \includegraphics[width=\linewidth]{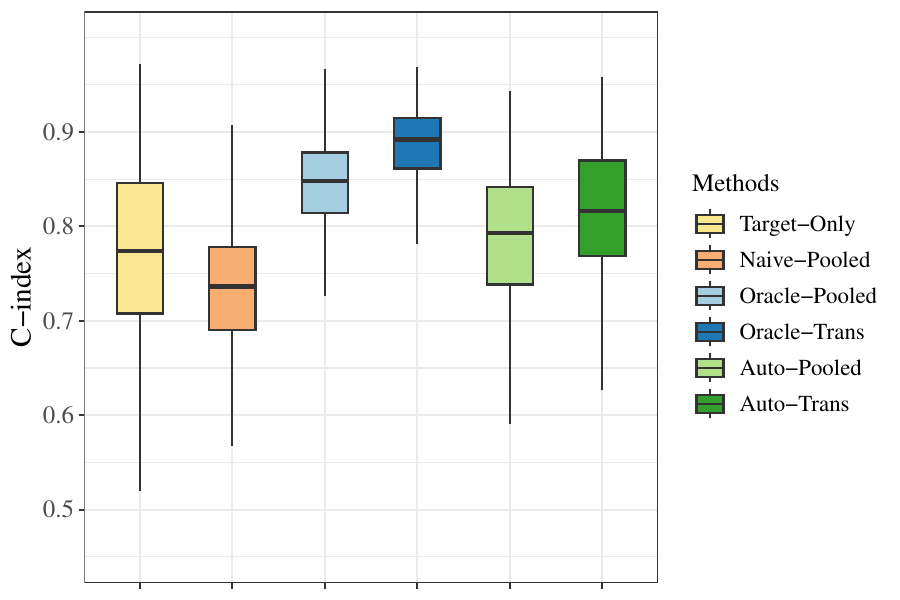}
         \caption{S5}
    \end{subfigure}
    \hfill
    \begin{subfigure}[b]{0.3\linewidth}
        \centering
        \includegraphics[width=\linewidth]{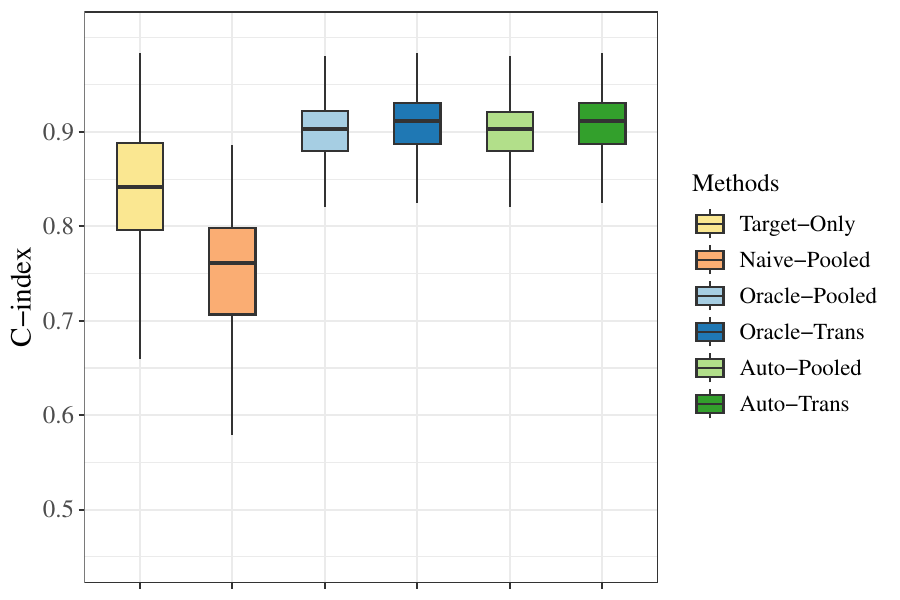}
         \caption{S6}
    \end{subfigure}
    \caption{Simulation results for C-index under Scenarios S1-S6}
    \label{fig:C-index}
\end{figure}

(2). From the results of the C-index presented in Figure \ref{fig:C-index}, we find that both Target-only and Naive-Pooled are inferior to the other four methods related to transfer learning. The Naive-Pooled, in particular, produces the smallest C-index. Among the four transfer learning methods, Oracle-Trans performs the best as expected, since it knows accurately and utilizes the useful sources. Auto-Trans also shows an absolute advantage and is comparable to Oracle-Trans, indicating that the proposed Algorithm \ref{alg:est_A} can accurately detect the informative sources. Besides, the debiasing step does help to improve the estimates since we observe that Oracle-Trans performs better than Oracle-Pooled and Auto-Trans performs better than Auto-Pooled. 

\begin{figure}[htbp]
    \centering
    \begin{subfigure}[b]{0.3\linewidth}
        \centering
        \includegraphics[width=\linewidth]{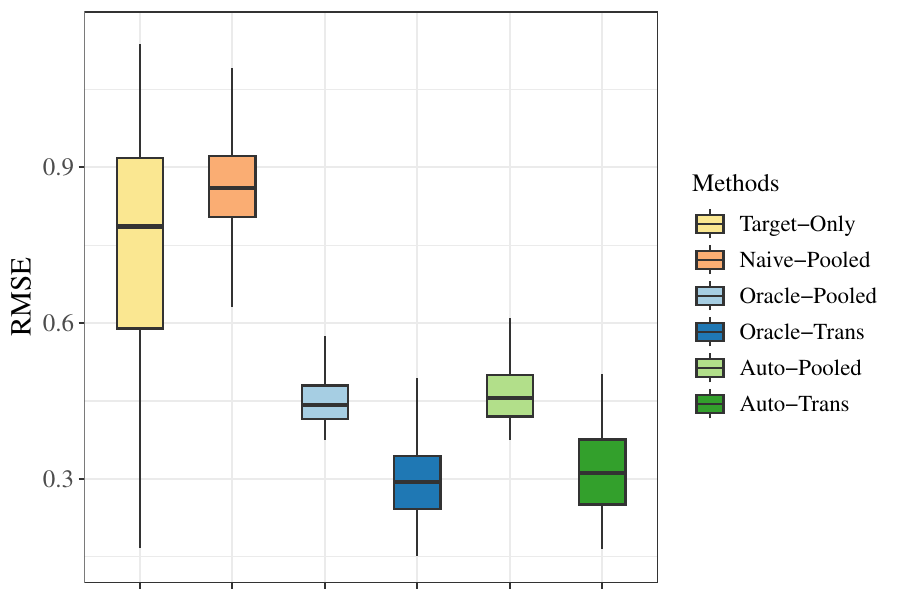}
         \caption{S1}
    \end{subfigure}
    \hfill
    \begin{subfigure}[b]{0.3\linewidth}
        \centering
        \includegraphics[width=\linewidth]{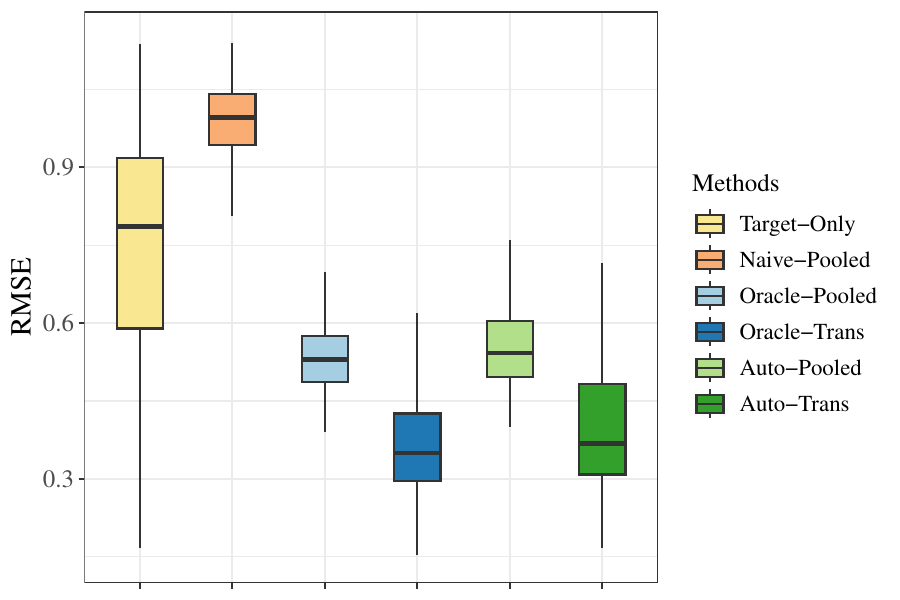}
         \caption{S2}
    \end{subfigure}
    \hfill
    \begin{subfigure}[b]{0.3\linewidth}
        \centering
        \includegraphics[width=\linewidth]{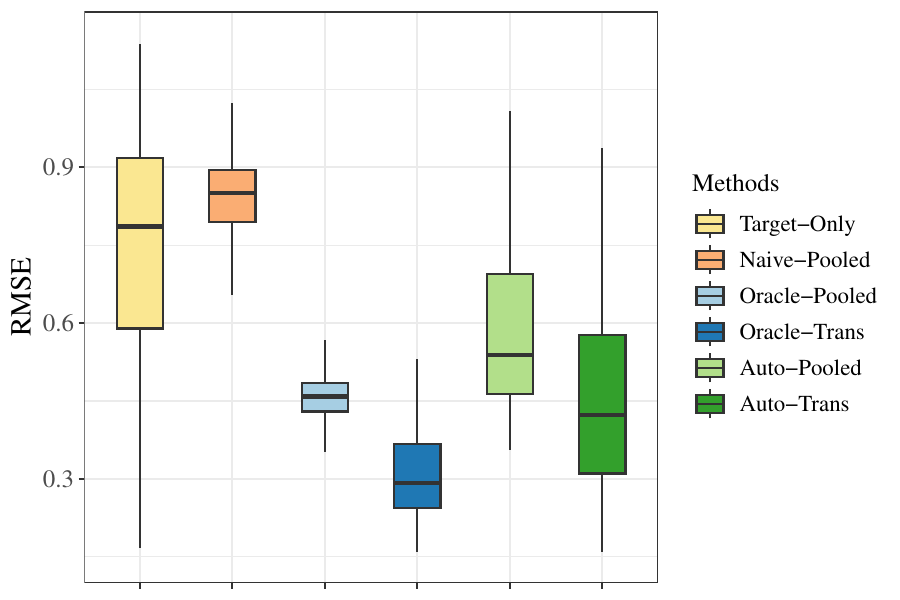}
        \caption{S3}
    \end{subfigure}

     \vspace{\floatsep} 
     
    \begin{subfigure}[b]{0.3\linewidth}
        \centering
        \includegraphics[width=\linewidth]{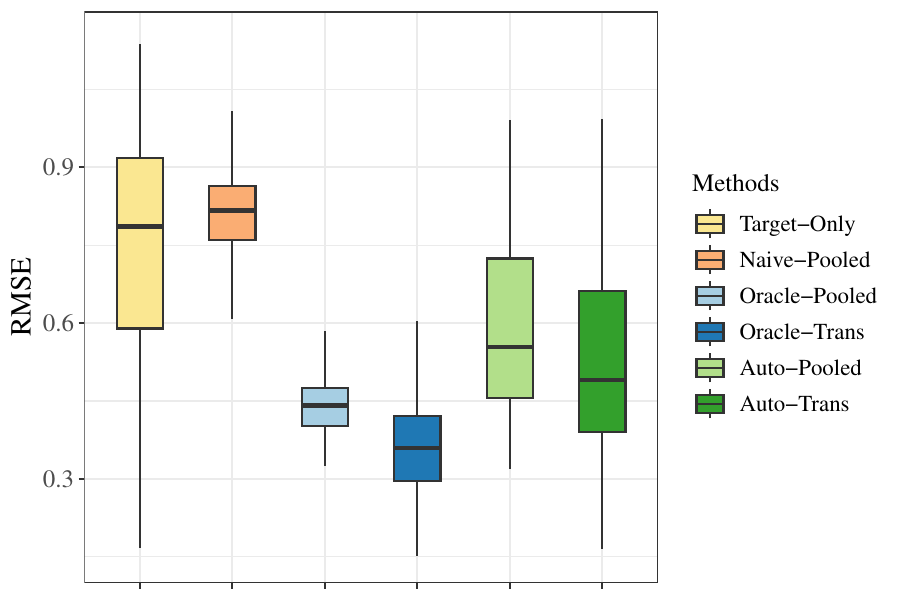}
         \caption{S4}
    \end{subfigure}
    \hfill
    \begin{subfigure}[b]{0.3\linewidth}
        \centering
        \includegraphics[width=\linewidth]{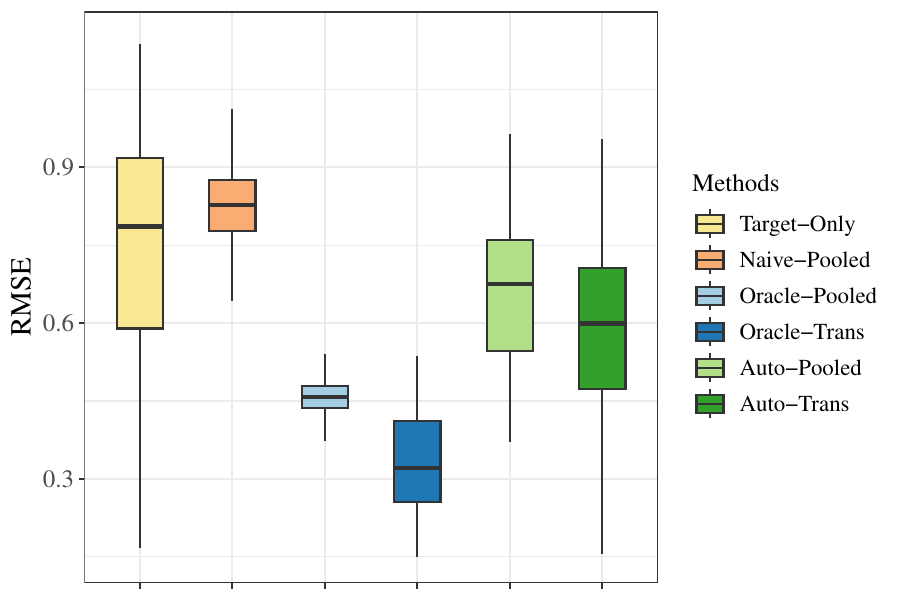}
         \caption{S5}
    \end{subfigure}
    \hfill
    \begin{subfigure}[b]{0.3\linewidth}
        \centering
        \includegraphics[width=\linewidth]{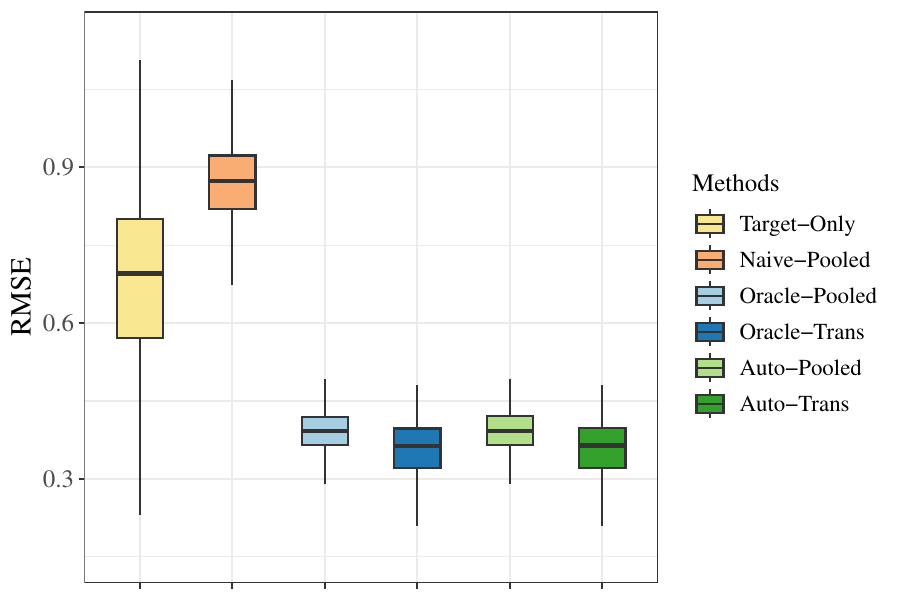}
         \caption{S6}
    \end{subfigure}
    \caption{Simulation results for RMSE under Scenarios S1-S6}
    \label{fig:RMSE}
\end{figure}

(3). {Figure \ref{fig:RMSE} shows the comparison of RMSE among different methods}. We can see that Oracle-Trans produces the smallest RMSE, as it accurately utilizes useful data sources, and Auto-Trans has a performance closer to Oracle-Trans in terms of RMSE, which confirms the superiority of the approach of detection for helpful sources. Besides, the transfer learning based methods have a clear advantage over Target-only and Naive-Pooled in estimation accuracy, since Target-Only ignores the helpful sources while Naive-Pooled incorporates many noisy sources. In addition, Oracle-Trans exhibits superior estimation accuracy compared to Oracle-Pooled, empirically confirming the importance of the debiasing step.

(4). By comparing the results across different scenarios, we find that the more similar the source and target data are, the more helpful it is to improve the estimates. Comparing the results in S2 and S3, we can see that the transfer learning methods have much better performances in terms of C-index and F1-score when the number of informative sources increased from one to three. The results from S3 and S4 show that there is a decrease in variable selection, estimation, as well as prediction accuracy as the sample size $n_k$ drops from 100 to 60 for the transfer learning methods. The performance of transfer learning methods in S5 is slightly worse than in S3 since the proportion of helpful sources is smaller. In the high-dimensional scenario S6, where $n_0$ is much smaller than $p$, the results show that the Fabs algorithm remains effective and the proposed methods also demonstrate superior performances.

(5). {Due to the space limit, we put the results of S7 in Supplementary Materials.} It is observed that as the number of helpful source datasets increases, the C-index gradually increases while the RMSE decreases. Similar to the results in scenarios S1-S6, Oracle-Trans outperforms Oracle-Pooled, while Auto-Trans outperforms Auto-Pooled, further illustrating the benefits of using target data for debiasing. Furthermore, as the number of helpful source datasets increases, the gap between Auto-Trans and Oracle-Trans narrows. This indicates that our proposed method is capable of attaining a performance comparable to that of the Oracle estimator when the number of informative source datasets is considerable. Besides, we calculate the Recall for the informative source dataset identification process, and the results show that the methods can accurately identify the informative sources with large Recall values. 


(6). {
The simulation results in Supplementary Materials for constructing confidence intervals show that the bias of the desparsified Lasso estimator is significantly reduced. In a large number of repeated simulations, the confidence intervals constructed by the proposed algorithm cover the true parameter value approximately $95\%$ of the time. Additionally, we plot the histograms of the desparsified Lasso estimator $\wt\bbeta$. The empirical distributions are in high agreement with the standard normal density, providing strong numerical evidence for the asymptotic normality established in Theorem \ref{th3}.}

\section{Real data analysis}
\label{section-4}
\subsection{Data preparation}

In this section, we focus on the sepsis cohort within the MIMIC-IV database (\url{https://mimic.mit.edu}), 
a contemporary electronic health record dataset spanning admissions from 2008 to 2019 \citep{johnson2023mimic}. Our research targets patients admitted to the ICU for the first time with sepsis. As mentioned in the introduction, MSSA is a highly significant and potentially lethal bacteria, yet current research on MSSA is insufficient and limited by inadequate data. To address this gap, we aim to utilize transfer learning tools for a more comprehensive analysis. Here the MSSA patient data serve as the target dataset, and other sepsis cases as source datasets. Sepsis cases with fewer than 40 samples and those of unspecified etiology are excluded, resulting in a total of nine source datasets, as shown in \Cref{tab:source_desc}. \Cref{tab:source_desc} demonstrates the ICD-code, sample size, and corresponding sepsis type for the nine source datasets. It also shows the gains in C-index for each source dataset using our proposed detection algorithm in \Cref{alg:est_A}, and seven source datasets with positive gains in C-index can improve the learning of the target model, while the other two with negative gains in C-index are not helpful.

The sepsis dataset to be analyzed includes one target dataset and nine source datasets, with a total sample size of 1700. Among these, 347 patients died of sepsis during the study periods leading to a censoring rate of $79.58\%$. We denote $T$ as the period from ICU admission to in-hospital death. For those who survive until hospital discharge, the survival time is censored with $C$ being the gap time between the discharge and the ICU admission dates. For each patient, 279 features are recorded but with a high missing rate, and variables with a missing rate over $25\%$ are discarded in our analysis. For continuous variables with a missing rate less than $10\%$, mean imputation is employed, and variables with a missing rate $10\%-25\%$ undergo multiple imputations. Missing values in binary categorical variables are filled in using the mode.  Ultimately, 102 covariates are included, categorized into: i) Demographic characteristics: age at admission, gender, and ethnicity; ii) Haematological assessments on the first day of ICU admission: haemoglobin, platelet, and white blood cell counts, electrolyte balance, renal and liver function tests, and coagulation profiles; iii) Arterial blood gas evaluations performed on the first day of ICU admission: lactate, pH, and oxygen saturation; iv) Scores measuring organ failure and illness severity: LODS, SOFA,  APS III, etc.


\subsection{Results}

\begin{table}[htbp]
  \centering
  \caption{Description of the target and source datasets}
  \label{tab:source_desc}
      \resizebox{\columnwidth}{!}{
    \begin{tabular}{cccccc}
    \toprule
    
    \multicolumn{1}{c}{Dataset} & \multicolumn{1}{c}{ICD-10} &  \multicolumn{1}{c}{Description} & Sample size & Gains in C-index & Selection times \\
    \midrule
    \multicolumn{1}{c}{Target} & \multicolumn{1}{c}{A4101} & Sepsis due to Methicillin susceptible Staphylococcus aureus & 229   & /     & / \\
    \midrule
     S1    & A4102  & Sepsis due to Methicillin resistant Staphylococcus aureus & 98    & -0.046 & 40 \\
    S2    & A408  & Other streptococcal sepsis & 79    & 0.003 & 69 \\
    S3    & A411  & Sepsis due to other specified staphylococcus & 74    & 0.001 & 89 \\
    S4    & A4151 & Sepsis due to Escherichia coli [E. coli] & 420   & -0.036 & 22 \\
    S5    & A4181 & Sepsis due to Enterococcus & 207   & 0.042 & 80 \\
    S6    & A4152  & Sepsis due to Pseudomonas & 88    & 0.006 & 66 \\
    S7    & A4189 & Other specified sepsis & 270   & 0.041 & 96 \\
    S8    & A4150  & Gram-negative sepsis, unspecified & 64    & 0.027 & 98 \\
    S9    & A4159  & Other Gram-negative sepsis & 171   & 0.028 & 83 \\
    \bottomrule
    \end{tabular}%
  }
\end{table}%

We use Target-Only, Naive-Pooled, and Auto-Trans methods for estimation due to the unknown true informative sources. The feature extraction results are shown in \Cref{tab:real_variableselect}. Our Auto-Trans method identifies 12 features, and Target-Only and Naive-Pooled identify 11 and 17 features, respectively. 
All three methods select admission age, Sequential Organ Failure Assessment score (SOFA), logistic organ dysfunction system (LODS) score, and three hematological assessment (first ICU day) indexes. Specifically, the estimated coefficients for admission age are negative in all three methods, indicating that older patients have shorter survival times and a higher risk of death. Additionally, SOFA is commonly used as a measure of organ dysfunction and has a high discriminative ability for predicting emergency and in-hospital mortality \citep{sofa2021}, which is consistent with the negative coefficient estimates. It is worth noting that Auto-Trans identifies the minimum total carbon dioxide level ($\mathrm{TCO_2}$) in blood gases, which is missed by both Target-Only and Naive-Pooled. Literature suggests that this feature is important, as low $\mathrm{TCO_2}$ levels are associated with higher risks of all-cause mortality \citep{yang2023associations}, which is consistent with the negative estimation coefficient. 


\begin{table}[htbp]
  \centering
  \caption{Data analysis: results of variable selection}
  \label{tab:real_variableselect}
    \resizebox{\columnwidth}{!}{
    \begin{tabular}{lccc}
    \toprule
    \multicolumn{1}{c}{Variable} & Target-Only & Naive-Pooled & Auto-Trans \\
    \midrule
    Admission age & \cellcolor[rgb]{ .867,  .922,  .969}-0.609 & \cellcolor[rgb]{ .988,  .894,  .839}-0.481 & \cellcolor[rgb]{ .886,  .937,  .855}-0.611 \\
    Maximum Partial Thromboplastin Time (PTT) recorded & \cellcolor[rgb]{ .867,  .922,  .969}-0.406 & \cellcolor[rgb]{ .988,  .894,  .839}-0.160 & \cellcolor[rgb]{ .886,  .937,  .855}-0.122 \\
    Logistic Organ Dysfunction System (LODS) & \cellcolor[rgb]{ .867,  .922,  .969}-0.305 & \cellcolor[rgb]{ .988,  .894,  .839}-0.480 & \cellcolor[rgb]{ .886,  .937,  .855}-0.489 \\
    Minimum anion gap recorded & \cellcolor[rgb]{ .867,  .922,  .969}-0.203 & \cellcolor[rgb]{ .988,  .894,  .839}-0.160 & \cellcolor[rgb]{ .886,  .937,  .855}-0.122 \\
    Minimum Partial Thromboplastin Time (PTT) recorded & \cellcolor[rgb]{ .867,  .922,  .969}-0.203 & \cellcolor[rgb]{ .988,  .894,  .839}-0.160 & \cellcolor[rgb]{ .886,  .937,  .855}-0.244 \\
    Sequential Organ Failure Assessment score (SOFA) & \cellcolor[rgb]{ .867,  .922,  .969}-0.102 & \cellcolor[rgb]{ .988,  .894,  .839}-0.160 & \cellcolor[rgb]{ .886,  .937,  .855}-0.244 \\
    Acute Physiology Score III & \cellcolor[rgb]{ .867,  .922,  .969}-0.203 & \cellcolor[rgb]{ .988,  .894,  .839}-0.160 & \textbackslash{} \\
    Minimum blood lactate level recorded & \textbackslash{} & \cellcolor[rgb]{ .988,  .894,  .839}-0.320 & \cellcolor[rgb]{ .886,  .937,  .855}-0.366 \\
    Verbal response score from the Glasgow Coma Scale (GCS) & \textbackslash{} & \cellcolor[rgb]{ .988,  .894,  .839}0.320 & \cellcolor[rgb]{ .886,  .937,  .855}0.122 \\
    Minimum heart rate recorded in vital signs & \textbackslash{} & \cellcolor[rgb]{ .988,  .894,  .839}-0.160 & \cellcolor[rgb]{ .886,  .937,  .855}-0.244 \\
    Mean respiratory rate recorded in vital signs & \textbackslash{} & \cellcolor[rgb]{ .988,  .894,  .839}-0.160 & \cellcolor[rgb]{ .886,  .937,  .855}-0.122 \\
    Simplified Acute Physiology Score II (SAPS II) & \textbackslash{} & \cellcolor[rgb]{ .988,  .894,  .839}-0.160 & \cellcolor[rgb]{ .886,  .937,  .855}-0.122 \\
    Minimum hematocrit level recorded & \cellcolor[rgb]{ .867,  .922,  .969}-0.102 & \textbackslash{} & \textbackslash{} \\
    Motor response score from the Glasgow Coma Scale (GCS) & \cellcolor[rgb]{ .867,  .922,  .969}0.203 & \textbackslash{} & \textbackslash{} \\
    Maximum partial pressure of oxygen (PaO2) in blood gases recorded & \cellcolor[rgb]{ .867,  .922,  .969}0.305 & \textbackslash{} & \textbackslash{} \\
    Minimum systolic blood pressure recorded in vital signs & \cellcolor[rgb]{ .867,  .922,  .969}0.305 & \textbackslash{} & \textbackslash{} \\
    Maximum hemoglobin level recorded & \textbackslash{} & \cellcolor[rgb]{ .988,  .894,  .839}0.160 & \textbackslash{} \\
    Maximum blood glucose level recorded & \textbackslash{} & \cellcolor[rgb]{ .988,  .894,  .839}0.160 & \textbackslash{} \\
    Minimum absolute eosinophil count recorded & \textbackslash{} & \cellcolor[rgb]{ .988,  .894,  .839}0.160 & \textbackslash{} \\
    Minimum total carbon dioxide level in blood gases recorded. & \textbackslash{} & \textbackslash{} & \cellcolor[rgb]{ .886,  .937,  .855}-0.030 \\
    Mean peripheral capillary oxygen saturation (SpO2) recorded in vital signs & \textbackslash{} & \cellcolor[rgb]{ .988,  .894,  .839}0.160 & \textbackslash{} \\
    Maximum glucose level recorded in vital signs & \textbackslash{} & \cellcolor[rgb]{ .988,  .894,  .839}-0.160 & \textbackslash{} \\
    \bottomrule
    \end{tabular}%
}
\end{table}%

To gain further insights into the analysis results, we conduct a random splitting-based evaluation. Since most of the source datasets are informative for predicting the target, we compare the performances of the three methods by reducing the number of informative sources. We randomly select $20\%$ of the target data as the testing set, with the remaining data used for training. The model is trained on the training set, and the C-index and Log-rank values are calculated on the testing set to evaluate the performance of the methods. Larger values of the Log-rank statistic indicate more significant differences in the survival curves between the high- and low-risk groups \citep{Harrington1982}, which in turn suggests better predictive accuracy of the model. Here, the high-risk and low-risk differences are delineated by the median of the model's predicted values $\wh\bbeta^{\top}\X$. This process is repeated 100 times to evaluate the performance of the various methods. 

\begin{figure}
    \centering
    \begin{subfigure}[b]{0.3\textwidth}
        \centering
        \includegraphics[width=\linewidth]{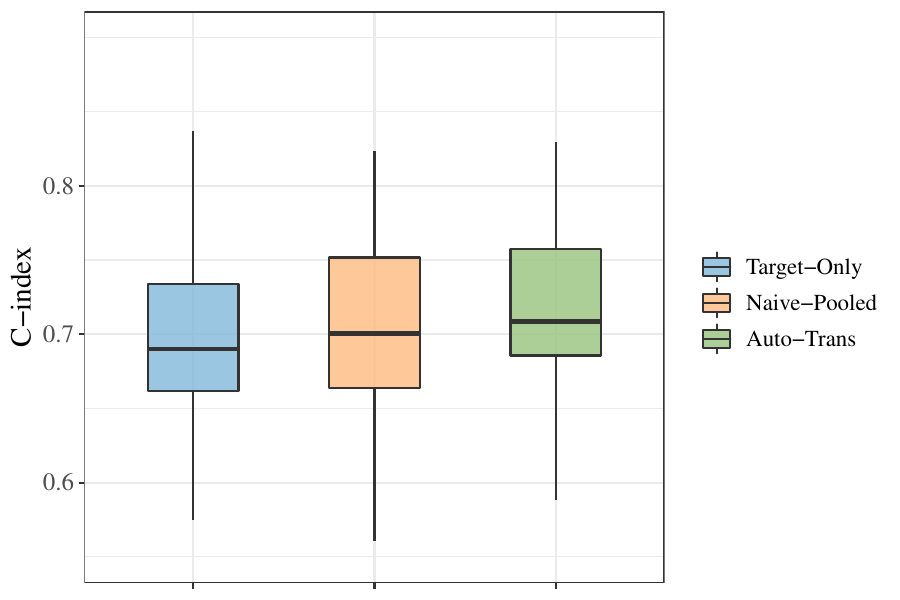}
         \caption{Scenario I: Three sources}
    \end{subfigure}
    \hfill
    \begin{subfigure}[b]{0.3\textwidth}
        \centering
        \includegraphics[width=\linewidth]{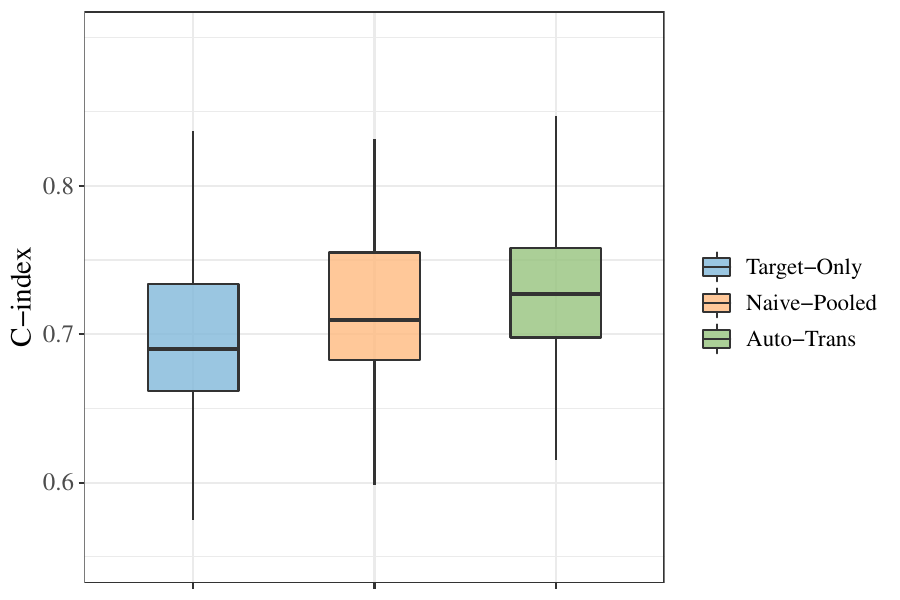}
        \caption{Scenario II: Six Sources}
    \end{subfigure}
    \hfill
    \begin{subfigure}[b]{0.3\textwidth}
        \centering
        \includegraphics[width=\linewidth]{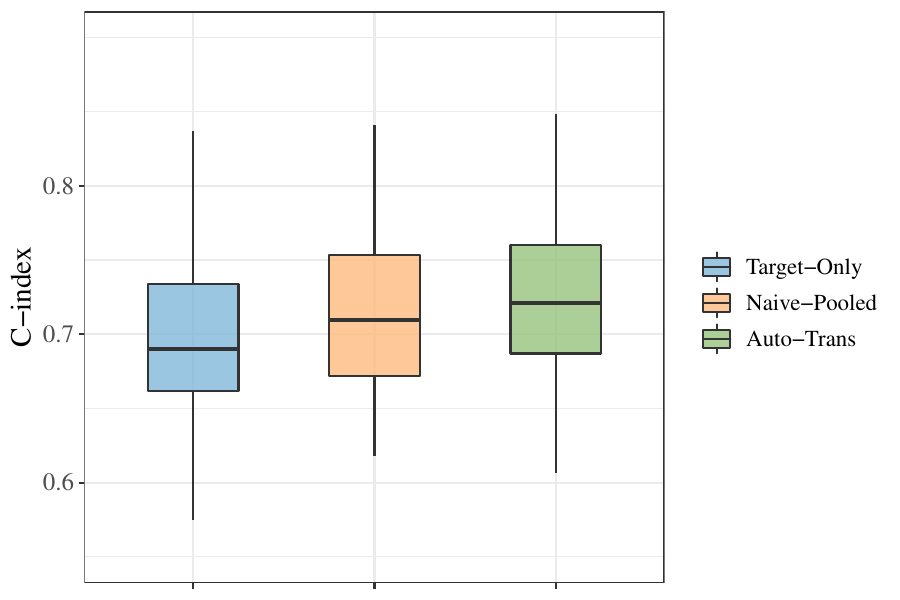}
        \caption{Scenario III: Nine Sources}
\end{subfigure}
      \vspace{\floatsep} 

      \begin{subfigure}[b]{0.3\textwidth}
        \centering
        \includegraphics[width=\linewidth]{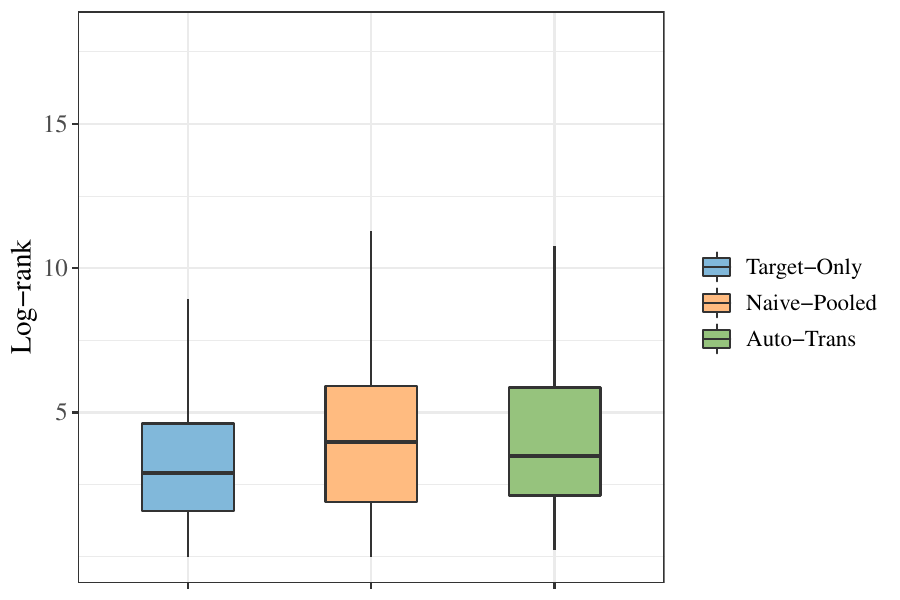}
         \caption{Scenario I: Three sources}
    \end{subfigure}
    \hfill
    \begin{subfigure}[b]{0.3\textwidth}
        \centering
        \includegraphics[width=\linewidth]{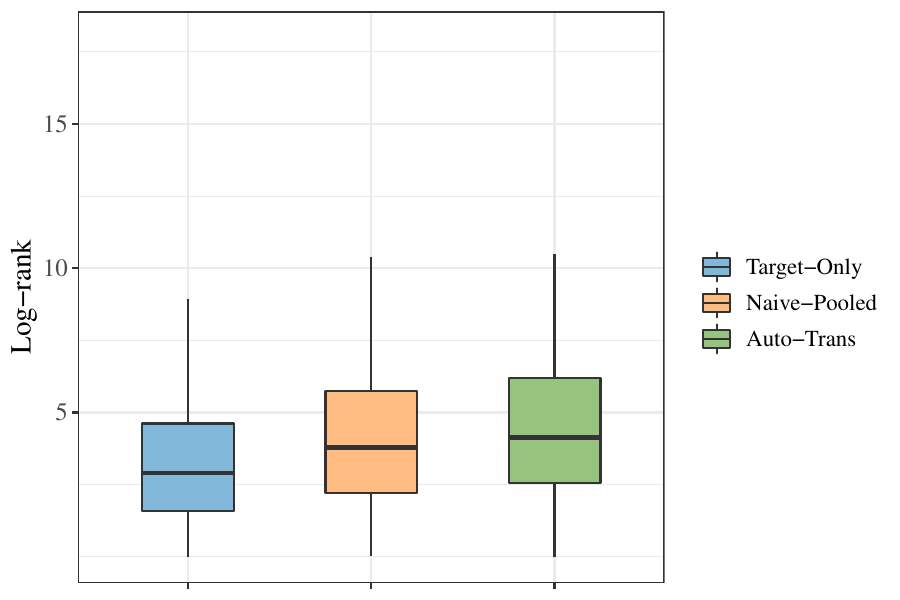}
        \caption{Scenario II: Six Sources}
    \end{subfigure}
    \hfill
    \begin{subfigure}[b]{0.3\textwidth}
        \centering
        \includegraphics[width=\linewidth]{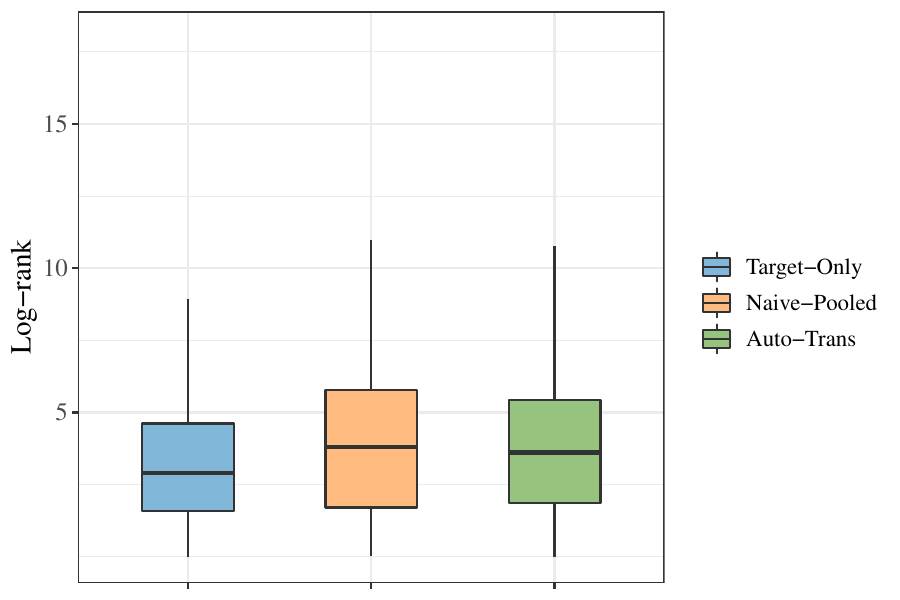}
        \caption{Scenario III: Nine Sources}
        \end{subfigure}
    \caption{Data analysis: results for random splitting-based evaluation}
    \label{fig:real}
\end{figure}

Among the nine sources, S1 and S4 are not helpful for promoting the target learning due to their negative gains in C-index, and we will keep them in the following evaluation process. We consider the following three scenarios: Scenario I pretends to have only three sources, S1, S4, and S8, where S8 has the largest selection times while S1 and S4 are the two sources with the smallest selection times (Table \ref{tab:source_desc}); In Scenario II, we consider six sources including S1, S2, S3, S4, S8, and S9; Scenario III includes all nine sources. Figure \ref{fig:real} shows the results based on 100 replications. 

The results show that Auto-Trans enjoys advantages in prediction accuracy over Naive-Pooled and Target-Only, since it tends to produce the largest C-index and Log-rank statistics. Target-Only has the poorest prediction performances due to the insufficient sample size. As the number of sources increases, the performance of Auto-Trans and Naive-Pooled improve in prediction accuracy as the proportion of helpful sources increases. Specifically, the mean C-index values for Auto-Trans in Scenarios I, II, and III are 0.717, 0.720, and 0.722 respectively, which are larger than the mean values (0.702,0.714,0.713) for Naive-Pooled, and 0.697 for Target-Only. For the three scenarios, the mean values of Log-rank statistics are (4.490,4.572,4.306) for Auto-Trans, (4.218,4.156,3.996) for Naive-Pooled, and 3.371 for Target-Only. Overall, clear advantages of transfer learning are again observed.

\section{Conclusion}
\label{section-5}

Transfer learning is a powerful tool to enhance model performance on the target dataset by leveraging information from other source datasets with similar but not exactly the same distributions. In this study, we have proposed a transfer learning approach (Auto-Trans) to address the pressing challenge of analyzing high-dimensional time-to-event data in the context of sepsis caused by MSSA. Given the complexity and high-dimensional nature of sepsis data, Auto-Trans, designed for transformation models, offers the flexibility of semiparametric models by capturing the relationship between survival time and predictors without relying on restrictive parametric assumptions. 
A key innovation of our method is the development of a transferable source detection mechanism based on the C-index, which can consistently identify informative sources and ensure that valuable information from related datasets is appropriately integrated. { Statistical validity is rigorously established. Furthermore, the confidence intervals for each coefficient component are provided with theoretical guarantees.}
Simulation results have shown that the proposed approach demonstrates competitive performance in enhancing variable selection, estimation, and prediction accuracy. When applied to the sepsis dataset, our method reveals findings that differ from alternative approaches, reaffirming its practical superiority.

Overall, our research contributes to the growing body of literature on transfer learning in survival analysis, offering a robust and scalable solution for high-dimensional time-to-event data with right-censored. Future work may explore the application of Auto-Trans to other data settings, such as truncated data and interval-censored data. Additionally, when several source datasets are available for analysis, it can be important and challenging to recognize the helpful source datasets to avoid negative transfer. A natural question is whether there is a technique that can avoid this effect of misidentification and it would be helpful to develop such methods that can adaptively use the non-informative sources without worrying about the negative transfer. 
{ Besides, \cite{li2024estimation} and \cite{he2024transfusion} recently proposed a new framework that jointly incorporates losses from both the target and source domains. In the scenario considered in this paper, applying this new framework yields the objective function:
\begin{eqnarray*}
\wt{Q}_n^{\Ah} & = 
-\sum_{k\in \Ah} \frac{\alpha_{k}}{n_{k}(n_{k}-1)} \sum_{i\neq j } \Delta_{i}^{(k)} I\left(Y_{j}^{(k)}>Y_{i}^{(k)}\right) S_{n}\left((X_{j}^{(k)}-X_{i}^{(k)})^\top\left(\delta^{(k)}+\beta^{(0)}\right)\right) \\
& \left. -\frac{\alpha_{0}}{n_{0}(n_{0}-1)} \sum_{i\neq j } \Delta_{i}^{(0)} I\left(Y_{j}^{(0)}>Y_{i}^{(0)}\right) S_{n}\left(X_{j}^{(0)^\top} \beta^{(0)}-X_{i}^{(0)^\top} \beta^{(0)}\right) \right. \\
& + \lambda_0 \left\|\beta^{(0)}\right\|_{1} + \sum_{k\in \Ah} 
\lambda_k \left\|\delta^{(k)}\right\|_{1}.
\end{eqnarray*}
We have compared the numerical performance of our proposed approach against this new framework through simulations in Supplementary material. However, the theoretical understanding in the differences between the two frameworks are complicated, and we leave this topic for future work.}



\section*{Acknowledgements}
 Lin's work was supported by the MOE Project of the Key Research Institute of Humanities and Social Sciences (22JJD910001).





\bibliographystyle{biom}
\bibliography{biomsample}

\label{lastpage}

\end{document}